\documentclass[useAMS,usenatbib]{mn2e}
\usepackage{amsfonts}
\usepackage{amsmath}
\usepackage[dvips]{graphicx}
\usepackage{epsfig}
\usepackage{wrapfig}
\usepackage{amssymb}
\usepackage{longtable}
\usepackage{accents}
\usepackage{dcolumn}
\usepackage{bm}
\usepackage[caption=false]{subfig}
\usepackage[normalem]{ulem}

\begin{document}

\title[A spinning supermassive black hole binary in S5 1928+738]{A spinning
supermassive black hole binary model consistent with VLBI observations of the S5 1928+738 jet}
\author[E. Kun, K. \'{E}. Gab\'{a}nyi, M. Karouzos, S. Britzen, L. \'{A}.
Gergely]{E. Kun$^{1,2}$\thanks{%
E-mail: kun@titan.physx.u-szeged.hu}, K. \'{E}. Gab\'{a}nyi$^{1,3}$, M.
Karouzos$^{4}$, S. Britzen$^{5}$, L. \'{A}. Gergely$^{1,2,6}$ \\
$^{1}$Department of Experimental Physics, University of Szeged, D\'om t\'er 9, H-6720 Szeged, Hungary\\
$^{2}$Department of Theoretical Physics, University of Szeged, Tisza Lajos krt 84-86, H-6720 Szeged, Hungary\\
$^{3}$Konkoly Observatory, MTA Research Centre for Astronomy and Earth Sciences, P.O. Box 67, H-1525 Budapest, Hungary\\
$^{4}$CEOU - Astronomy Program, Department of Physics \& Astronomy, Seoul National University, Gwanak-gu, 151-742, Seoul, Korea\\
$^{5}$Max-Planck-Institute f\"{u}r Radioastronomie, Auf dem H\"{u}gel 69, D-53121 Bonn, Germany\\
$^{6}$Department of Physics, Faculty of Science, Tokyo University of Science, 1-3, Kagurazaka, Shinjuku, Tokyo 162-8601, Japan}
\date{Accepted . Received ; in original form }
\maketitle

\begin{abstract}
Very Long Baseline Interferometry (VLBI) allows for high-resolution and high-sensitivity observations of
relativistic jets, that can reveal periodicities of several years in their
structure. We perform an analysis of long-term VLBI data of the quasar S5
1928+738 in terms of a geometric model of a helical structure
projected onto the plane of the sky. We monitor the direction of the jet axis
through its inclination and position angles. 
We decompose the variation of the inclination of
the inner $2$ milliarcseconds of the jet of S5 1928+738 into a periodic term
with amplitude of $\sim 0.89^{\circ }$ and a linear decreasing trend with
rate of $\sim 0.05^{\circ }\text{yr}^{-1}$. We also decompose the variation of the position
angle into a periodic term with amplitude of $\sim 3.39^{\circ }$ and a
linear increasing trend with rate of $\sim 0.24^{\circ }\text{yr}^{-1}$.
We interpret the periodic components as arising
from the orbital motion of a binary black hole inspiraling at the jet base and derive corrected values of the mass ratio and separation from the accumulated 18 years of VLBI data.
Then we identify the linear trends in the variations as due to the slow reorientation of the
spin of the jet emitter black hole induced by the spin-orbit precession and we determine the precession period $T_{\mathrm{SO}}=4852 \pm 646$ yr of the more massive
black hole, acting as the jet emitter. Our study provides indications, for the first time from VLBI jet kinematics, for the spinning nature of the jet-emitting black hole.
\end{abstract}

\pagerange{\pageref{firstpage}--\pageref{lastpage}} \pubyear{2014}

\label{firstpage}


\section{Introduction}

The existence of Supermassive Black Holes (SMBH) at the centre of galaxies, combined with the important role of
galaxy mergers in the hierarchical galaxy formation models and the long time-scale of the SMBH merging process
suggest that SMBHs often exist in pairs 
\citep[][]{Blandford1986iau,SearleZinn1978,Komossa2006}. The binary passes through three
evolutionary stages during the galaxy merger \citep{Begelman1980}. At first
the stellar halo of each galaxy interacts with the central black hole of the
other galaxy via dynamical friction \citep[see
e.g.][]{BinneyTremaine1987}. Gradually the black holes sink into a common
gravitational potential well, and a bound, widely separated SMBH binary forms 
\citep[a detailed review of massive black hole binary
evolution can be found in][]{MerrittMilos2005}. After this, a transition stage follows. The separation then
decreases due to dynamical friction and beyond a transition radius the
dissipation of energy and angular momentum becomes dominated by gravitational
radiation. General relativistic effects during this inspiral stage are
described in a post-Newtonian (PN) framework by expanding the field equations in terms of the PN
parameter $\varepsilon =Gmc^{-2}r^{-1}$ (where $m$ is the total mass, $r$ is the separation, $G$ is the gravitational constant, and $c$ is the speed of light). PN techniques
are involved in the description of merging binary-dynamics if $0.001<\varepsilon<0.1$, while in the plunge and the ring-down stages
numerical calculations are required to characterize the final SMBH. The
spins of the black holes generate corrections to the dynamics at $1.5$PN
orders through the spin-orbit coupling, at $2$PN through the spin-spin
coupling and at even higher orders through various post-Newtonian
corrections to these couplings. The complicated PN dynamics in the presence
of the spins is described in detail in e.g. \citet{Barker1975,Barker1979}, \citet{Kidder1995}, \citet{Gergely2010b,Gergely2010a}. When the
spin and orbital momentum vectors are not aligned, a spin-orbit type
precession will cause the spin vectors to slowly rotate about the orbital
angular momentum vector. 

SMBHs are believed to be the main engine of the activity in Active
Galactic Nuclei (AGN) \citep[for a review see e.g.][]{Begelman1984}. Apart from efforts of identifying binary black hole candidates in kpc-scale separated binaries with spatially
resolved AGN \citep[e.g.][]{Komossa2003}, indirect approaches have been also employed for the cases when gravitational radiation already
dominates the merger of a sub-pc-scale separated binary.
It has been estimated from the jet
power, that at least one merger with orbital period of the order of one year
should be detected in future blind surveys (at a nominal sky coverage of $
10^{4}$ deg$^{2}$ with $0.5$ mJy sensitivity) through the electromagnetic (EM)
signature of the SMBH binary on the jet \citep{Shaughnessy2011}, manifesting
itself as an increase in the jet power when the binary inspirals toward its
barycentre.

Relativistic bulk motion in the jet of AGN causes Doppler boosting of the
synchrotron radiation of the relativistic particles. Very Long Baseline
Interferometry (VLBI) observations reveal superluminally outward moving
objects in a number of relativistic jets. Such superluminal motions are only
apparent, indicating projection effects. These sources often show extreme
variability throughout the EM spectrum. The radio variability can be
explained in terms of flaring events following the ejection of jet-components or of a wiggling jet, implying a variable Doppler factor.
 AGN variability is however mostly non-periodic, hence detecting
periodic behaviour of the radio light curve and precessing jets would
strongly indicate a SMBH binary at the jet base 
\citep[e.g. OJ287,][]{Villata1998}. While a plausible explanation for the precession of the
jet is the existence of a merging SMBH binary at its base, alternative
models allowing for a single black hole and external causes for the jet
precession have been proposed \citep[e.g. see the Discussion in][and references therein]{Britzen2010}. 

The precession of a jet originating in a SMBH binary system was discussed in e.g.
\citet[][]{Roos1993,Romero2000,Britzen2001,Lobanov2005,Valtonen2012,Caproni2012}.
Several models have been developed to investigate the interaction between a
SMBH binary and its astrophysical environment, revealing the presence of the
binary in the AGN \citep[e.g.][]{MacFadyen2008,Tanaka2013}.
Recently, the helical modulation of the large-scale radio jet of J1502+1115 was attributed to the effect of the tight pair in a triple supermassive black hole system \citep{Deane2014}.
Observational
data have already been employed to identify precessing jet
candidates. AGN observables allowed placing constraints on some of the SMBH binary
parameters, like the total mass, the orbital period and the separation of
the binary, e.g. Mrk 501, \citet{Villata1999}; 3C 273, \citet{Romero2000}; BL Lac,  \citet{Stirling2003}; 3C 120,  \citet{Caproni2004}; 3C 345, \citet{Lobanov2005}; S5 1803+784, \citet{Roland2008}; NGC 4151, \citet{Bon2012}.
The effect of the spin was not included in the calculations of these works, and the binary motion
was approximated by pure Keplerian orbits.

The extreme Kerr-limit of the rotating SMBH is characterized by the value $1$ of the dimensionless spin parameter $\chi$ (the spin $S$ can be calculated as $S= G c^{-1} m^2 \chi$). However, as
the accretion disk radiates and some of that radiation is trapped by the
black hole, the extreme limit is actually not reached and rather the
so-called canonical spin limit of $\chi _{\mathrm{can}}\simeq 0.998$ applies \citep{Thorne1974}. A closed magnetic field line topology further connects
the black hole horizon to the accretion disk, field lines acting as anchor
chains, which further reduces the spin to $\chi \simeq 0.89$ \citep{Kovacs2011}. Nevertheless this value is still extremely high, and as
the presence of the jet signals a rich black hole environment, supporting
the assumption of accretion, it is plausible to assume that at least the jet
emitter black hole spins fast, and as such, the spin effects in the binary
dynamics are non-negligible. The spin-powered jets are believed to be aligned with the black hole spin axis \citep{Blandford1977}, therefore precessing jets can reveal the presence of the spin.

In this paper we investigate the source S5 1928+738. This is a
core-dominated quasar  at a redshift $z=0.302$, its luminosity distance is $D_{\mathrm{L}}=1620$ Mpc, and at this distance the spatial resolution is  $4.6$ pc/milliarcsecond 
\citep[where
the cosmological parameters are $\Omega _{\mathrm{m}}=0.314$, $\Omega _{\lambda}=0.686$, $H_{0}=67.4$ km s$^{-1}$ Mpc$^{-1}$,][]{Planck2013}. The source is a well
studied member of a complete sample of the extragalactic flat spectrum radio
sources taken from the S5 strong source survey \citep{Kuhr1981}. The VLA map
at a wavelength of $20$ cm \citep{Johnston1987} reveals a structure comprising
two lobes and a bright radio-core. Zooming into the jet base a one-sided core-jet structure appears on the milliarcsecond (mas) scale. The maximal apparent velocity
detected in the jet, $8.1c$ \citep[][]{Lister2013} suggests
high-speed bulk motion that is nearly head on, the mean jet direction is
pointing close to the observer line of sight (LOS). This feature makes the jet a strongly
Doppler-boosted one.

An analysis of the arcsecond- and the mas-scale jet structure has been
published by \citet{Hummel1992} and the jet structure has been found to exhibit
wiggles on both scales. A moving sine wave was fitted to the temporal
motion of the components in the jet. \citet{Roos1993} proposed that the host
galaxy of S5 1928+738 harbours a binary SMBH at its centre, inducing the
periodic structures detected in the jet. A total mass of $10^{8}$ M$_{\odot}
$ was adopted, which would account for the bolometric luminosity on the
basis of the Eddington limit. The orbital period was calculated to be $\sim 2.9$
years, the orbital separation $\sim 0.005$ pc and the mass ratio $>0.1$. It has also been noted that if the
wiggles in the mas-scale jet of S5 1928+738 are
due to the orbital motion of a massive binary, the mean direction of the jet
should obey a Lense-Thirring precession \citep{LenseThirring1918} about the orbital
angular momentum of the binary, with a period of the order of $10^{3}$ yr.
\citet{Murphy2009} used the VLBI Space Observatory Programme data to further constrain the binary parameters by fitting their equations to jet component data from $8$ epochs of observation performed at 5 GHz. The derived parameters are however affected by their assumption of the jet being
emitted by the fictitious reduced mass particle.

The data collected on S5 1928+738 within the framework of the Monitoring Of Jets in Active galactic nuclei with VLBA Experiments (MOJAVE) survey
span almost twenty years \citep{Lister2009}. Careful analysis of this dataset allows us to detect
and quantitatively analyse for the first time the spin-orbit precession of
the jet emitter SMBH, manifesting itself as a small change over time in the
direction of the jet axis. The spatial-length of the jet based on the MOJAVE-VLBI maps at $15$ GHz and assuming an inclination of $7\degr$ is approximately $560$ pc.

The paper is organised as follows. In Section 2 we summarize the
observational techniques which can reveal periodicities in the
extragalactic jets observed at radio wavelengths. More specifically we
present methods based on the investigation of the AGN jet morphology, AGN
jet kinematics and AGN jet flux density variability.
In Section 3 we present the analysis of the archival MOJAVE data of S5 1928+738, investigating the temporal
evolution of the jet structure. In Section 4, we describe a geometrical model for the three-dimensional (3D)
helical jet structure. By projecting the spatial jet onto the plane of the sky and employing the maximal
apparent velocity seen in the jet (which provides information about the inclination angle), we constrain the
model by fitting it to the MOJAVE observations on the jet. Based on the jet modelling and on the independent total mass estimate, we give the binary parameters in Section 5. In
Section 6 we discuss our findings and give our conclusions.

\section{Identifying jet precession through observations}
\label{Identifying jet precession through observations}

Three main techniques have been employed to identify and study AGN jet
precession.

\subsection{AGN jet morphology}
\label{AGN jet morphology}

VLBI observations offer unparalleled spatial resolution and flux sensitivity
compared to other wavelength regimes. VLBI studies thus allow the detailed
investigation of the morphology of AGN jets and the inference of both
physical and geometrical properties.\newline
AGN jets have been known to exhibit strong curvature along their length and
at all scales ranging from sub-pc to Mpc, although curvature appears to
increase closer to the VLBI core 
\citep[e.g.][]{Krichbaum1994,Krichbaum1995, Alberdi1997,Walker2001,
Caproni2004, Savolainen2006}. The curvature is usually identified in terms
of the apparent position of individual jet components \citep[e.g.][]{Britzen2008}, but also in terms of the jet as a whole. This
is accomplished by the definition of the jet ridge line, a line that
connects the centres of all VLBI jet components identified at a single epoch
of observations \citep[e.g.][]{Hummel1992,Britzen2010,Karouzos2012a}. The
jet ridge line of AGN jets can be used to identify jets with strong
variations of their curvature along the jet axis. \citet{Karouzos2012b},
using a simple geometrical model of a helical jet, showed that a helically
structured or a precessing jet can give rise to the curvature observed in
many AGN jets.

\subsection{AGN jet kinematics}
\label{AGN jet kinematics}

VLBI observations at different epochs can reveal not only the overall shape
of the jet but also the position and proper motion of the individual
components that make up the jet. This proper motion is characteristic of
both the physics of the components themselves \citep[e.g.][]{Marscher1985}
and the properties of the underlying jet, namely the presence of precession
at the foot-point of the jet. In particular, non-ballistic motions, i.e.,
acceleration along the components' trajectories and/or significant
non-radial motions can reveal the presence of a precessing jet. It has been
shown that many AGN jet components follow helical trajectories which can be
explained by means of plasma instabilities 
\citep[e.g. Kelvin-Helmholtz
instability,][]{Camenzind1992, Hardee1997, Perucho2006}.\newline
Several well-studied AGN have been found to show non-ballistic motions in
their jets, which are interpreted in terms of a helical or precessing jet 
\citep[e.g.][]{Steffen1995,Abraham1999,Britzen2009,Liu2010,Kudryavtseva2011a,Karouzos2012b}.\newline
The projected positions of individual components on the plane of the sky can
be directly used for fitting of analytical models and can therefore usually
constrain both the opening angle of the precession cone, the initial time
and angle of ejection of each jet component, and given enough data, the
periodicity of the jet precession and consequently of the binary black hole
system. As described in the previous section, the sensitivity of the VLBI
observations together with the density of the temporal sampling determine how
accurately the precession properties can be derived. In addition, the study
of individual jet component kinematics entails the cross-identification of
components across epochs.

\subsection{Flux density variability}
\label{Flux density variability}

Long-term monitoring of a source in the optical and/or radio allows the
construction of a light curve that can be analysed in terms of jet
precession and the presence of a BBH system. Given the relativistic speed in
AGN jets, small variations in the angle between the trajectory of a jet
component and our line of sight (LOS) can lead to significant flux
variations due to the Doppler effect. Such angle changes are predicted in a
precessing jet. In addition, for a precessing jet (unlike, e.g. for an
expanding shock wave within the AGN jet) flaring events at different
wavelengths are expected to be synchronous rather than lagging \citep[e.g.][]{Pyatunina2007,Kudryavtseva2011b}. By analysing periodicities
in AGN light curves \citep[e.g.][]{Fan2007,Kudryavtseva2011a} and more
importantly by analysing the cross-correlation between light curves at
different wavelengths \citep[e.g.][]{Quirrenbach1991,Britzen2009}, one can
deduce the periodicity of a possible precession. This methodology has been
used to derive the parameters of a BBH system in OJ 287 \citep{Villata1998},
and in 3C 454.3 \citep{Qian2007}.\newline
The difficulty of such flux variability studies lies in the dense sampling
of the light curve that is required to robustly extract periodicity time-scales. In addition, the time-scales usually predicted by models of jet
precession and BBH systems are of the order of several years \citep[e.g.][]{Roland1994,Steffen1997,Caproni2004}, implying that a source
must be monitored for at least a similar (or longer) time period.\newline
\newline
For the few sources for which both kinematic and photometric information is
available, a combination of the above three methods can constrain any
precession or binary black hole model further. Notable examples include S5
1803+784 \citep{Roland2008}, PKS 0420-014 \citep{Britzen2001}, and 3C 345 \citep{Lobanov2005}. In the next section
we present the radio analysis of the mas-scale jet of S5 1928+738.

\section{Analysis of archival MOJAVE data}
\label{Analysis of archival MOJAVE data}
\subsection{Archival VLBA data}

Observations of S5 1928+738 used in the present work were taken between 1994.67 and 2013.06 in $45$ epochs
with the Very Long Baseline Array (VLBA) at 15 GHz within the framework of the
MOJAVE survey. The VLBA
comprises ten radio telescopes, each with a diameter of $25$ m. The
longest baseline corresponds to $8611$ km and the maximal number of possible
baselines is $45$. The half-power beam width is $0.47$ mas at $2$ cm ($15.36$ GHz). For more details of the
observational setups see \citet{Lister2013} and references therein.
We downloaded the $45$ epochs of calibrated visibilities of S5~1928+738 from the MOJAVE webpage\footnote{http://www.physics.purdue.edu/astro/MOJAVE/} and
model-fitted them using the Caltech \textsc{Difmap} software package \citep{Shepherd1997}. The image parameters and main results concerning our model-fits are summarized in Table \ref{table_impars}. During the model-fitting process circular or elliptical Gaussian components are used to parametrize the brightness
distribution of the jet and model the jet as the sum of a number of 2-dimensional Gaussian components. 
Performing model-fitting on data of several years one can trace the temporal change in the
properties of the components; any change reflects the evolution of the jet.

We also made use of the 43 GHz VLBA map of the jet from \citet{Lister2000} obtained at 1999 January 13, that is close to the MOJAVE epoch of 1999 January 2. This high frequency archival data provided better resolved jet structure compared to the 15 GHz data.

\begin{figure}
\begin{center}
\includegraphics[scale=0.4]{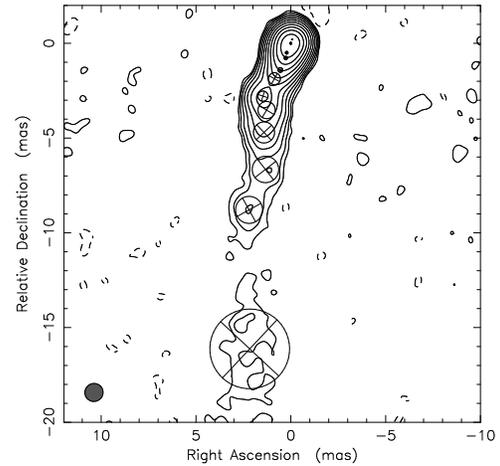}
\end{center}
\caption{Model-fits to the 2013.06 epoch of MOJAVE observation of S5 1928+738 at $15$ GHz. The core lies at $0,0$ coordinates. Contours are in percent of the peak flux $2.18$ Jy beam$^{-1}$ and they increase by factors of two. The lowest contour level is $1.1$ mJy beam$^{-1}$ ($0.05$ percent of the peak flux). The off-source rms noise is $0.16$ mJy beam$^{-1}$. The restoring beam-size is shown in the bottom left corner of the image ($0.9 \mathrm{mas} \times 0.9 \mathrm{mas}$). Each circle on the map corresponds to a model-fit component. Their sizes represent the FWHM of the fitted
Gaussian. Tapered image was created in \textsc{Difmap}.}
\label{map}
\end{figure}

\begin{figure*}
\begin{centering}
\includegraphics[scale=0.38,angle=270]{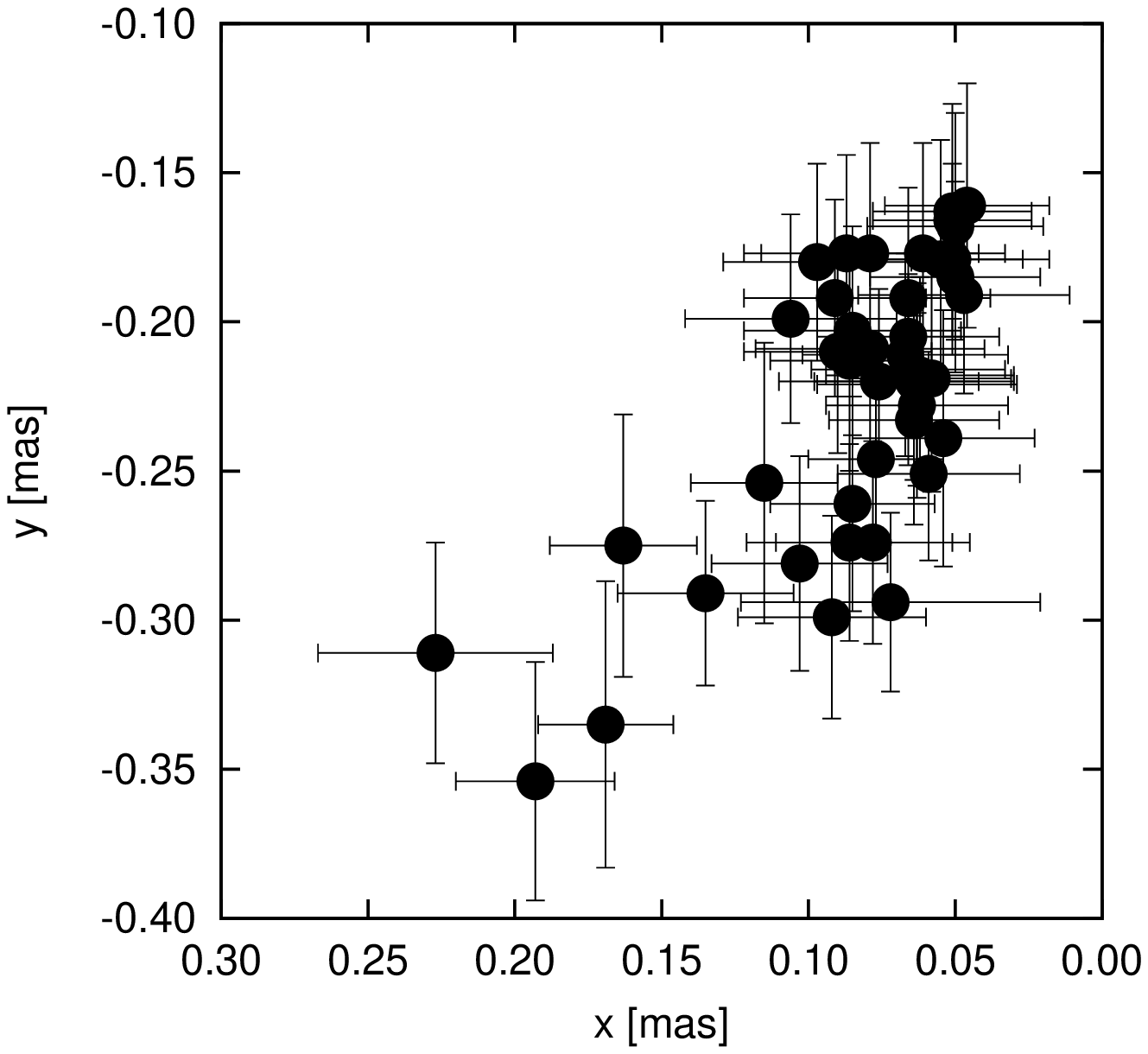}
\hspace{0.5cm}
\includegraphics[scale=0.38,angle=270]{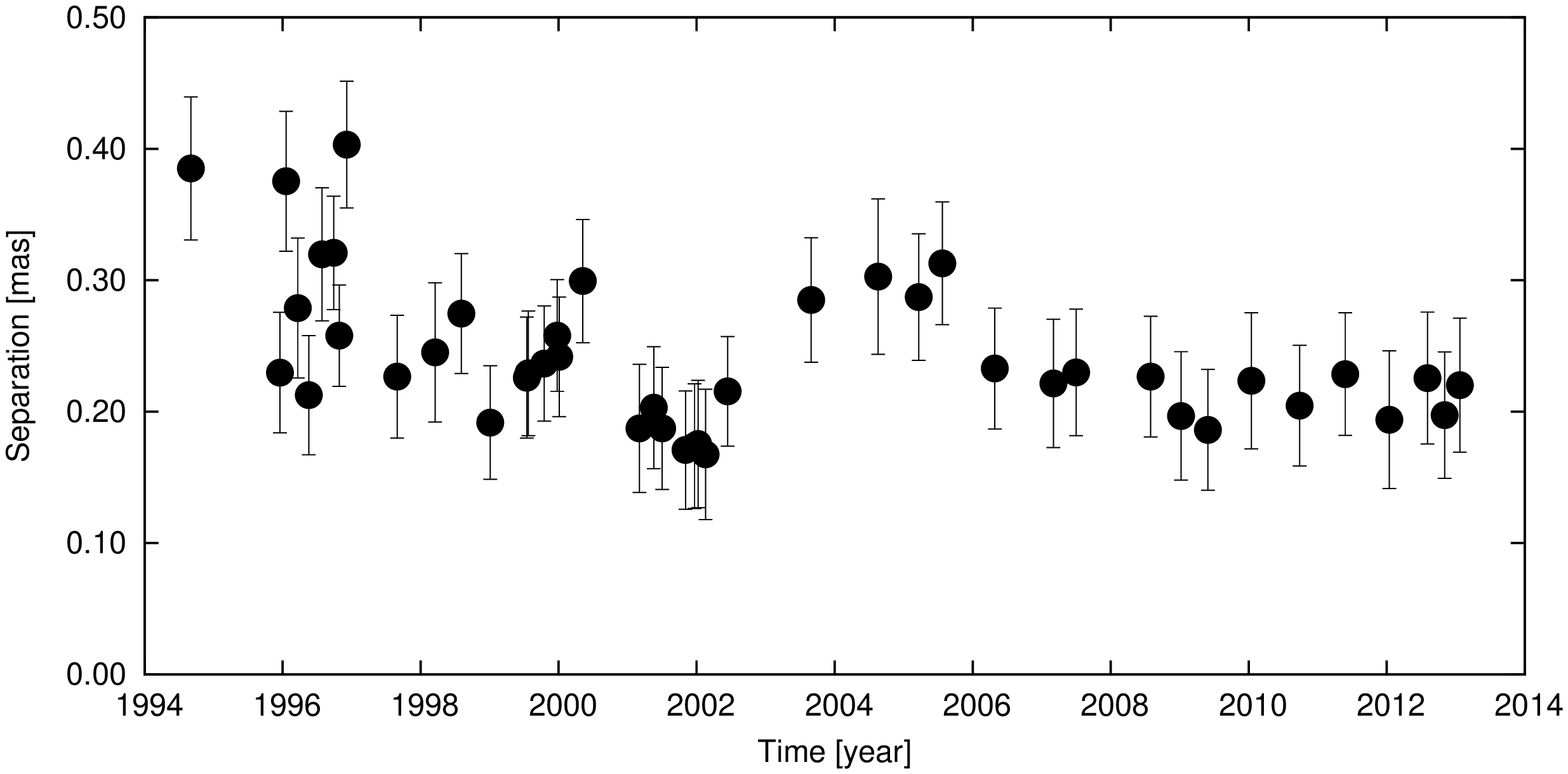}
\caption{\textit{Left}: Relative position of model-fit jet component Cg with respect to the position of model-fit jet component CS across the epochs at $15$ GHz measured in the system, where East defines the x-axis and the North the y-axis. Cg appears to be stationary within the errors, except for $5$ epochs implying the five most Eastern positions (1994.67, 1996.05, 1996.57, 1996.74, 1996.93). \textit{Right}: Relative separation of Cg with respect to CS plotted against time.}
\label{cgmotion}
\includegraphics[scale=0.45,angle=270]{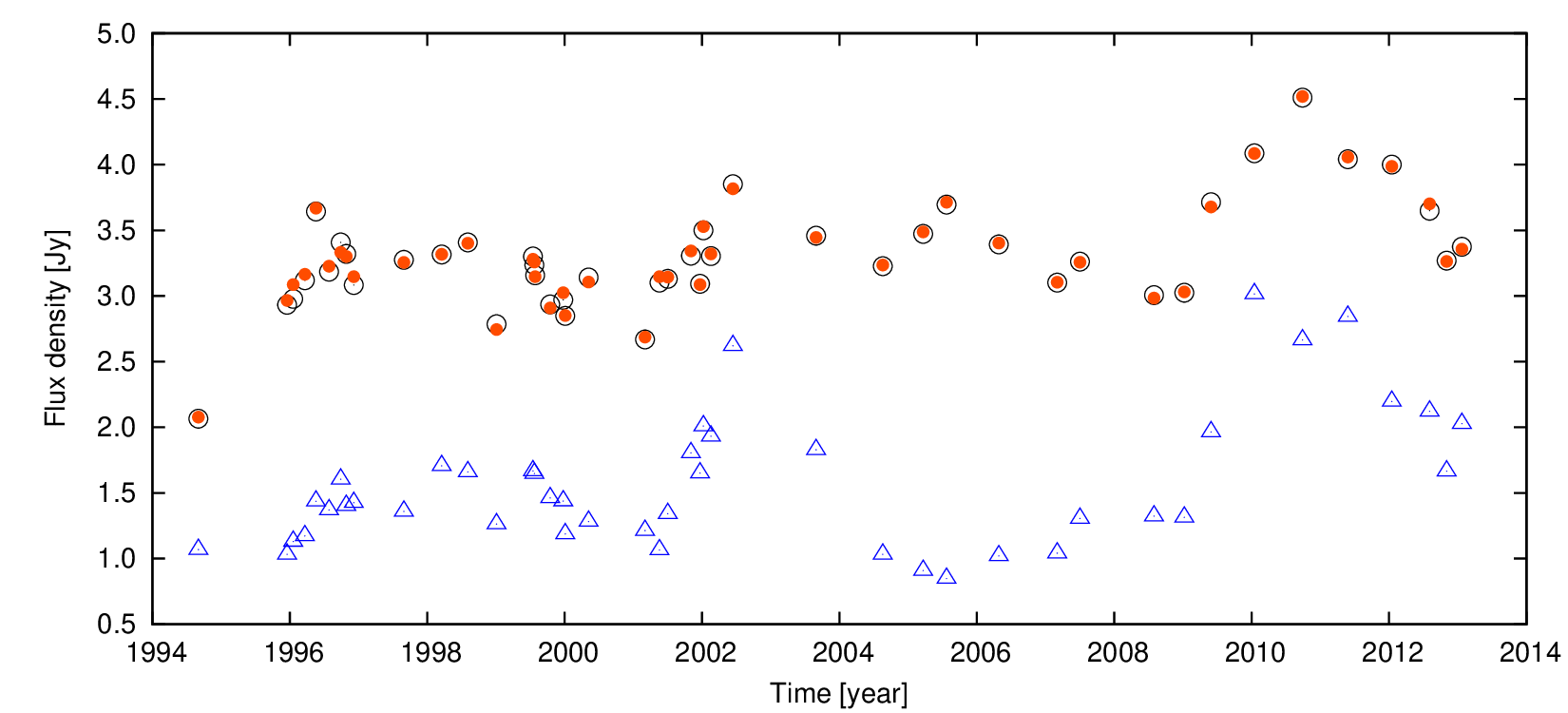}
\caption{Flux density of the core region (combined flux densities of CS and Cg, blue triangle), total flux density of the jet measured by the MOJAVE (black open circle), total flux density of the components in the recent work (orange filled circle) plotted against time.}
\label{comp_flux}
\end{centering}
\end{figure*}

\begin{table*}
\begin{minipage}{140mm}
\begin{center}
\caption{Summary of the 15 GHz image parameters. (1) epoch of the VLBA
observation, (2) VLBA experiment code, (3)--(4) FWHM minor and major axis of the restoring
beam, respectively, (5) position angle of the
major axis of the restoring beam measured from North through East, (6) rms noise of the image, (7) reduced $\chi{^2}$ of the \textsc{Difmap} model-fit, (8) number of the components in the model. The full table is available in electronic format online.}
\centering
\begin{tabular}{cccccccc}
\hline
\hline
	&	VLBA  & $\mathrm{B}_{\mathrm{min}}$ & $\mathrm{B}_{\mathrm{maj}}$& $\mathrm{B}_{\mathrm{PA}}$ & rms  & Red. & Comp.\\
Epoch & Code & [mas] & [mas] & [$^{\circ}$] & mJy bm$^{-1}$ & $\chi{^2}$& Number \\
(1) & (2) &  (3) &  (4) &  (5) &  (6) & (7) & (8)\\
\hline
1994 Aug 31	&	BZ004	&	0.67	&	0.76	&	-67	&	0.4	&	1.21 & 7\\
1995 Dec 15	&	BK37A	&	0.53	&	0.76	&	12	&	0.7	&	0.39 & 11\\
1996 Jan 19	&	BR034	&	0.44	&	1.00	&	-9	&	0.6	&	1.16 & 11\\
1996 Mar 22	&	BR034B	&	0.46	&	1.08	&	-16	&	0.4	&	1.01 & 11\\
1996 May 16	&	BK037B	&	0.52	&	0.83	&	-39	&	0.3	&	1.39 & 11\\
1996 Jul 27	&	BR034D	&	0.48	&	0.97	&	-16	&	0.4	&	2.55 & 10\\
1996 Sep 27	&	BR034E	&	0.47	&	0.91	&	-43	&	0.4	&	1.19 & 8\\
1996 Oct 27	&	BK037D	&	0.42	&	0.65	&	22	&	0.5	&	0.85 & 11\\
\hline
\hline
\label{table_impars}
\end{tabular}
\end{center}
\end{minipage}
\end{table*}
\begin{table*}
\begin{center}
\caption{Circular Gaussian model-fit results for S5 1928+738. (1) epoch of observation, (2) flux density, (3)--(4) position of the component center with respect to the core, (5) FWHM, (6) jet-component identification. The full table is available in electronic format online.}
\begin{tabular}{lccccc}
\hline
\hline
Epoch [yr] & Flux density [Jy] & x [mas] & y [mas] & d [mas]  & CO\\ 
\hline
1994.67 & 0.606 $\pm$ 0.044 & 0.000 $\pm$ 0.040 & 0.000 $\pm$ 0.037 & 0.145 $\pm$ 0.003 & CS\\   
        & 0.462 $\pm$ 0.045 & 0.227 $\pm$ 0.040 & -0.311 $\pm$ 0.037 & 0.144 $\pm$ 0.005 & Cg \\
        & 0.407 $\pm$ 0.042 & 0.561 $\pm$ 0.048 & -1.238 $\pm$ 0.045 & 0.301 $\pm$ 0.007 & C7 \\
        & 0.408 $\pm$ 0.042 & 0.258 $\pm$ 0.051 & -1.613 $\pm$ 0.049 & 0.350 $\pm$ 0.007 & C6 \\
        & 0.078 $\pm$ 0.020 & 1.153 $\pm$ 0.087 & -3.282 $\pm$ 0.086 & 0.789 $\pm$ 0.064 & C3 \\
        & 0.038 $\pm$ 0.022 & 2.203 $\pm$ 0.206 & -7.417 $\pm$ 0.206 & 2.029 $\pm$ 0.580 &  C2\\
        & 0.035 $\pm$ 0.012 & 2.675 $\pm$ 0.142 & -10.212 $\pm$ 0.141 & 1.366 $\pm$ 0.175 & C1 \\
\hline
\hline
\label{longtable}
\end{tabular}
\end{center}
\end{table*}
\subsection{Error estimation and component identification}
\label{error_compid}

To decrease the degrees of freedom during the model-fit and to remain
consistent through the epochs we used only circular Gaussian components to fit
the brightness profile of the jet. In the analysis of the VLBA
data by the MOJAVE group the core was fitted at several epochs by one elliptical
component with axial ratio $<0.5$. We fitted the core region by two
circular components in each epoch 
without increasing the reduced $\chi^2$ with
respect to the MOJAVE analysis given in \citet{Lister2009}. We also checked the higher frequency data of \citet{Lister2000} from 1999 January 13 to clarify the presence of the
northernmost component and concluded that it is visible at $43$ GHz as
well. The northernmost component is marked with 'CS' and the other jet-component in the core region is labelled by 'Cg' in accordance with the high frequency analysis of \citet{Lister2000}.

The fitted parameters were the position ($x$[mas], $y$[mas]), the integrated flux
density ($S_{t}$ [Jy]) and the full width at half maximum (FWHM) ($d$[mas]) of the Gaussian
components. A representative image of the jet can be seen in Fig. \ref{map}. There is a prominent bend at an approximate core separation of $3$ mas. \citet{Hough2013} suggests that a similar bend seen in 3C 207 is related to the re-collimation zone in the jet of that source.

For each component the post-fit root mean square error ($\sigma _{p}$) was
determined on the residual image after the jet components were extracted from
the original map around the position of each component. The peak flux of the
components ($S_{p}$) was estimated from $S_{t}$ and $d$. With $\theta_{\mathrm{min}}$ and $\theta_{\mathrm{maj}}$ denoting the minor and major axis of the
restoring beam, the beam size is given as $\theta=\sqrt{\theta_\mathrm{min}^{2}+\theta_{\mathrm{maj}}^{2}}$. Then the error of the integrated flux density can
be given as \citep{Schinzel2011} 
\begin{flalign}
{\sigma}_t=\left({\sigma}_p \cdot \sqrt{1+SNR} \right) \sqrt{\left(1+\frac{S_t^2}{S_p^2}\right)},
\end{flalign}where $SNR=S_{p}/{\sigma }_{p}$. The error of the component width is 
\begin{flalign}
{\sigma}_d = 
\begin{cases}
{{\sigma}_p \cdot \sqrt{\theta ^2+d^2}}/{S_p}, & \theta>d
\\ 
{{\sigma}_p \cdot d}/{S_p}, & \theta\leq d.
\end{cases}
\end{flalign}
Concerning the estimation of the position error we followed the
suggestion of \citet{Lister2009} and assumed that the uncertainties on the
position of the fitted components are $\sim 10\%$ of the component size
convolved with the beam size. Then the position errors are given as: 
\begin{flalign}
& {\sigma}_x=0.1\cdot \sqrt{\theta _x^2+d^2},\\
& {\sigma}_y=0.1\cdot \sqrt{\theta _y^2+d^2},
\end{flalign}where $\theta_{x}$ and $\theta_{y}$ are beam projected
to the $x$ and $y$ axis, respectively. For our full sample containing even the faintest components, we estimate the average errors of the integrated flux densities, positions and widths to be $20$\%, $8$\% and $12$\%, respectively. 

The positions of the components can be most reliably determined for the most compact features. However, the larger the core separations, the larger the FWHM of the fitted components; therefore, the positional uncertainties increase as a function of the core distance. The identification of the outer jet shape is thus difficult.

Finally we
cross-identified components across different epochs by requiring a smooth temporal change for the separation with respect to the core, the flux
density, and the FWHM of the fitted Gaussian. The components are labelled with letter 'C' and a number, so that the larger number denotes the closest component to the core. The six jet components labelled with B1–-B6 each appear in one epoch only. They cannot be followed through several epochs in a similar way to the standalone components marked as C.

The MOJAVE survey does not consist of phase-referenced observations, therefore there is no absolute
positional information. To study the movement of the components, one has to choose a reference point. This
reference point is usually the so-called core, which is assumed to be stationary. In the case of our model-fit of S5 1928+738, we have chosen the northernmost component (denoted by CS) as the core. This choice will be revisited and confirmed in Section \ref{thebehaviourofthejet}. The parameters of the model-fit components are given in Table \ref{longtable}.

\subsection{The behaviour of the jet}
\label{thebehaviourofthejet}

In addition to the core, $10$-$12$ components can be identified at $15$ GHz
during the almost $20$ years of radio observations. Most of the
components are moving away from the core CS except Cg, which appears at nearly constant position with respect to the core (see Fig. \ref{cgmotion}).
 In the first few epochs, though, there is an apparent shift in its position. This can originate from projection effects, or observational effects. The latter possibility is related to the fact, that in these first few epochs the beam of the array was significantly different, more elliptical than in later epochs. 

We calculated the linear proper motion of the components. The maximal proper motion obtained in the jet is $\mu_{\mathrm{max}}=0.43 \pm 0.02$ mas yr$^{-1}$ that is in excellent agreement with the maximal proper motion $0.43 \pm 0.01$ mas yr$^{-1}$ by \citet{Lister2013}. We note, however, the fastest component in our analysis is C3 (at an average core distance $6$ mas), while the fastest component in the analysis of \citet{Lister2013} is the outermost component. Most probably the different choice of the reference point and the slightly different component identifications led to this difference between the two analysis. 

The standard theory of the superluminal motion yields a connection between the apparent speed ($\beta_\mathrm{app}$), the intrinsic jet speed ($\beta_\mathrm{jet}$), and the inclination angle ($\iota$): 
\begin{flalign}
\beta_\mathrm{app}\!=\!\frac{\beta_\mathrm{jet} \sin\iota}{(1-\beta_\mathrm{jet} \cos \iota)}.
\label{betaapp}
\end{flalign}
Taking the largest observed proper motion of $\mu_{\mathrm{max}}=0.43 \pm 0.02$ mas yr$^{-1}$, the maximum apparent velocity seen in the jet is $\beta_\mathrm{app}\approx 8.1$c and therefore the lower limit of the Lorentz factor is $\Gamma_\mathrm{min}\approx8.1$. The minimal intrinsic jet velocity is $\beta_{\mathrm{jet}}=0.992$c, and the corresponding critical inclination angle, at which the apparent velocity is maximum, is $\iota_\mathrm{{c,max}} \approx 7\degr$.


The other consequence of the Doppler boosting is that the apparent brightness temperature $T_\mathrm{b}$ can exceed the limiting intrinsic brightness temperature $T_\mathrm{int}$. In case of VLBI components the brightness temperature is calculated as \citep[e.g.][]{Condon1982}:
\begin{flalign}
T_\mathrm{b,VLBI}[\mathrm{K}]=1.22\cdot 10^{12} (1+z) \frac{S_{\nu}}{\theta^2 \nu^2} ,
\end{flalign}
where $S_{\nu}$[Jy] is the flux density, $\theta$[mas] is the FWHM of the component, and $\nu$[GHz] is the observing frequency. We calculated $T_\mathrm{b,VLBI}$ for the brightest component in those epochs, at which the FWHM of this component was larger than $0.05$ mas (roughly one tenth of the beam size). The apparent and intrinsic brightness temperatures are related to each other as $T_\mathrm{b}=\delta T_\mathrm{int}$, where $\delta=\Gamma^{-1}(1-\beta_\mathrm{jet} \cos \iota)^{-1}$ is the Doppler factor. We use $T_\mathrm{int}\approx10^{11}$K as the maximal limit on the particle temperature \citep{finnek1999}, at which the Compton catastrophe occurs. Thus the average Doppler factor becomes $\delta_{\mathrm{VLBI}}\approx8.5$. Using the values derived from the jet kinematics, $\Gamma=8.1$, $\beta=0.992$, and $\iota=7\degr$, the Doppler factor is estimated to be $\delta\approx8$, that is in good agreement with the Doppler factor derived from the brightness temperature of the components.

Considering this result and the large number of the observed jet-components having proper motion estimates in S5 1928+738, we expect that the bulk motion in the jet is not significantly faster than it is indicated by the observed maximum apparent velocity $8.1$c. We adopt the above value of the inclination angle as initial value in the jet-fitting presented in Section \ref{Applying the model on the jet of S5 1928+738}.

The component Cg is the brightest in $32$ out of $45$ epochs. The behaviour of the Cg component raises the question whether the northernmost component CS is indeed the core or if it should be Cg instead.
In the following, we discuss the components Cg and CS. If we assume that Cg is the core, then CS could be regarded as part of the counter-jet. In that case, assuming that the jet and counter-jet are symmetrical and
have an intrinsic jet velocity of $\beta=0.992$, an inclination angle of $\iota=7.0^\circ$, a flat spectral index of
$\alpha_{\mathrm{182MHz-8.4GHz}}=-0.05$ \citep{CRATES2007}, and a continuous jet with the parameter n=$2$ \citep{Scheuer1979}, we can calculate the jet counter-jet ratio as
\begin{flalign}
R\!=\!\left(\frac{1+\beta \cos\iota}{1-\beta\cos\iota}\right)^{(n +\alpha)}\approx 13000.
\end{flalign}
This implies that for CS to be part of the counter-jet and be observable it needs to be intrinsically $13000$ times brighter than the jet. As this is not physically plausible, we conclude that the CS is not part of the counter-jet, but rather is the core.

The flux densities of the core region, the total flux density of the jet measured by the MOJAVE and from our model-fit process are displayed in Fig. \ref{comp_flux}. It can be seen that the flux variation of the core region dominates the total flux density across the epochs. The flux density variations in the core region are suggestive of periodic behaviour, indicating a variable jet-source at the jet base. The jet components are expanding with increasing core separation, as expected from their adiabatic cooling that occurs as they move away from the core.

We note, that the sampling of the total flux density is nearly homogeneous except for the first epoch, when there is a time gap of $1.3$ year between the first and the second epochs. Because the flux curve is poorly determined between 1994-1996 and we can identify only $7$ components in the first epoch, we excluded the first epoch (1994.67) from our jet analysis presented in Section \ref{Applying the model on the jet of S5 1928+738}.

\section{Geometrical model for the jet}

In this section we will use a geometric model to describe the jet
morphology as developed by \citet{Nakamura2001}. They modelled the AGN radio jets by 3D magnetohydrodynamic (MHD) simulations based on the Sweeping Magnetic-Twist model. According to this a Poynting flux of torsional Alfv\'{e}n wave train produced in the interaction of the rotating accretion disk and the large scale magnetic field results in a slender jet shape by the sweeping pinch effect. Wiggled structures in the jets are produced by the helical kink instability, as the torsional Alfv\'{e}n wave train encounters a domain of reduced Alfv\'{e}n velocity, where the toroidal component of the field accumulates, resulting in a helical structure of the jet.

In what follows, we describe the mathematical formalism
used to fit the jet geometry (Section \ref{Revealing the jet shape}), and the application of this model to
the jet (Section \ref{Applying the model on the jet of S5 1928+738}).

\subsection{Revealing the shape of the jet}
\label{Revealing the jet shape}

Following \citet{Nakamura2001}, we model the jet of S5 1928+738 with a
helical structure and describe the jet pattern with a conical helix in the following way:
\begin{flalign}
& x_\mathrm{jet}\!=\!\frac{b}{2\pi }u\cos u, \nonumber \\
& y_\mathrm{jet}\!=\!\frac{b}{2\pi }u\sin u, \nonumber \\
& z_\mathrm{jet}\!=\!\frac{a}{2\pi }u,
\label{coniceqs}
\end{flalign}
where $u$ is the polar angle measured in the plane perpendicular to the jet
cone axis, and $a$ and $b$ are the axial and radial growth rates of the jet per turn, respectively. We use the following coordinate
system where the $x$ and $y$ axes are in the plane of the sky and
\begin{description}
\item $x$ axis points to East,
\item $y$ axis points to North,
\item $z$ axis is the direction of the LOS.
\end{description}
The projection of the jet onto the $xy$ plane can be carried out by two
rotational matrices. The $R_{+\iota{_0}}$ matrix
\begin{flalign}
R_{+\iota{_0}} \!=\!
\begin{pmatrix}
\cos \iota{_0} & 0 & \sin \iota{_0} \\ 
0 & 1 & 0 \\ 
-\sin \iota{_0} & 0 & \cos \iota{_0}
\end{pmatrix}
\label{riota}
\end{flalign}
rotates the structure around the $y$ axis from the LOS toward the plane of the
sky with $\iota_0$ angle, that is the inclination angle between the
LOS and the jet axis. The $0$ index refers to the angles of the jet
geometrical axis. If the jet geometrical axis lies in the plane of the sky, 
$\iota_0$ equals $90^\circ$, if the jet axis
points exactly to the observer, $\iota_0$ equals $0^\circ$. The $R_{+\lambda{_0}}$ matrix
\begin{flalign}
R_{+\lambda{_0}}\!=\!
\begin{pmatrix}
\cos \lambda{_0} & -\sin \lambda{_0} & 0 \\ 
\sin \lambda{_0} & \cos \lambda{_0} & 0 \\ 
0 & 0 & 1%
\end{pmatrix}%
\label{rlambda}
\end{flalign}
rotates the structure around the $z$ axis with $\lambda_0$, that is the 
position angle of the jet symmetry axis measured in the plane
of the sky with respect to the $y$ axis. Applying the rotational matrices on the
helix shape, the projected helical structure in 2D becomes: 
\begin{flalign}
\label{proj_geom1}
& {x_\mathrm{jet}^p}(u)\!=\!F(u)\cos(\lambda{_0})-G(u)\sin(\lambda{_0})\\
& {y_\mathrm{jet}^p}(u)\!=\!F(u)\sin(\lambda{_0})+G(u)\cos(\lambda{_0})
\label{proj_geom2}
\end{flalign}
where
\begin{flalign}
& F(u)\!=\!\frac{b}{2 \pi} u\cos( u-\phi)\cos(\iota{_0})+\frac{a}{2\pi}u\sin(\iota{_0})\\
& G(u)\!=\!\frac{b}{2\pi} u\sin(u-\phi)
\end{flalign}
and $\phi$ is an initial phase.\newline
The intrinsic half-opening angle of the cone ($\psi^\mathrm{int}$) can be given by the radial and axial increments as 
\begin{flalign}
\tan(\psi^\mathrm{int})\!=\!b/a.
\end{flalign}
The apparent half-opening angle of the cone ($\psi ^\mathrm{obs}$) is related to $\psi ^\mathrm{int}$ as
\begin{flalign}
\psi^\mathrm{obs}(0\degr<\iota _{0}<90\degr)\!=\!\frac{\psi ^\mathrm{int}}{\sin \iota_0 }.
\label{psziobs}
\end{flalign}

\subsection{Applying the model to the jet of S5 1928+738}
\label{Applying the model on the jet of S5 1928+738}

We deal with the above described helical jet shape and present the application of
the geometrical model. In the present context, we do not mark the components, but just assume that they
are part of the jet and reveal the jet shape in one epoch; we fit the jet as
a whole instead of fitting the individual components' path. 

The compact jet of S5 1928+738 extends over $15$ mas. Because roughly half of the jet-components lie within core separation of $2$ mas in every epoch and dominate the flux density, and because the inner jet (core separation $\leqq2$ mas) can be more precisely model-fitted (see Section \ref{error_compid}) than the outer jet (core
separation $>2$ mas), in the following we deal only with the inner part of jet. In addition to the above reasoning, any precession of the jet due to a binary would primarily affect the inner jet, with the effects of the precession being dampened as the core separation increases.

For resolved jets, there is evidence for non-zero opening angles. \citet{Perucho2012} analysed multi-epoch VLBI data of the radio jet in the quasar S5 0836+710 at
different frequencies and found that the jet opening angles are similar at all the used frequencies: $1.6$, $2$, $5$, $8$,
$15$, $22$, and $43$ GHz.

\begin{figure}
\begin{center}
\includegraphics[scale=0.3]{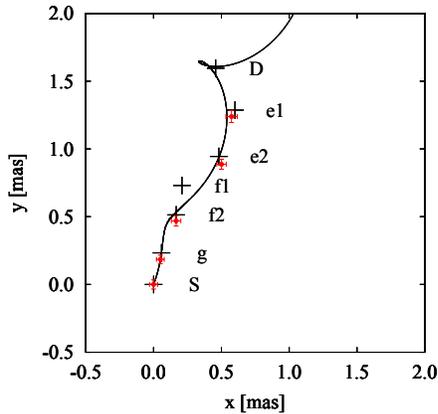}
\end{center}
\caption{Black crosses denote the 43 GHz model components derived by \citet{Lister2000} (their sizes are not representative of the errors), red dots with error bars denote the $15$ GHz model components. In the figure, North is oriented towards negative y values. We complement the data by five of the seven component positions identified at $15$ GHz. The black curve represents the ``starting jet model'' (see Section \ref{Applying the model on the jet of S5 1928+738}) fitted to the 43 GHz data. }
\label{startingmodel}
\end{figure}

In accordance with the finding that the opening angle does not depend on the typical VLBI observing frequencies \citep[e.g.][and references therein]{Perucho2012},
we use the $43$ GHz data to derive the intrinsic
half-opening angle of the jet as an input parameter for
modelling of the $15$ GHz datasets. The model-fit results for the inner $2$ mas of the jet at $43$ GHz data given by \citet{Lister2000} are in good agreement
with our model-fit results of the most closely separated 15 GHz data (see Fig. \ref{startingmodel}) but contain more components, describing the jet shape more precisely.
Thus we set up a starting model of the
helical jet by fitting Eq. (\ref{proj_geom1}) and Eq. (\ref{proj_geom2}) on the component-positions
of the inner jet measured at $43$ GHz. 

The degeneracy between the inclination angle and the intrinsic half-opening angle does
not allow for these parameters to be fitted together, see Eq. (\ref{psziobs}). Therefore, one needs to
 estimate one of these angles independently. In Section \ref{thebehaviourofthejet} we derived the inclination angle $\iota=7.0\degr$ from the apparent velocity seen in the jet. The method is independent from the geometrical model and therefore we fix the inclination of the jet axis $\iota_0=7.0\degr$ in the starting model at $43$ GHz.

\begin{table}
\begin{center}
\caption{Starting parameters for the geometrical model obtained by jet component positions at $43$ GHz \citep{Lister2000}. The inclination angle, $\iota_0$, was calculated from the maximum observed apparent velocity (see Section \ref{thebehaviourofthejet}).}
\begin{tabular}{@{}c c c c}
\hline
\hline
$\iota_0$[$\degr$] & $\lambda_0$[$\degr$] & $a$ [mas] & $b$ [mas]\\
\hline
$7.00$ & $160\pm 2$ & $10.6\pm 0.4$ & $0.18\pm0.06$\\
\hline
\hline
\end{tabular}
\label{pars_mastercone}
\end{center}
\end{table}

We fitted the 43 GHz components' positions with the conical helix described by Eq. (\ref{proj_geom1}) and Eq. (\ref{proj_geom2}). The resulting parameters of the fit are given in
Table \ref{pars_mastercone}. Based on our parametric fit, the jet parameter $\Gamma \psi^\mathrm{int}$ is $0.14\pm0.05$, using $\Gamma=8.1$ and $\psi^{\mathrm{int}}=\arctan(b/a)=0.017$ $\mathrm{rad}\pm0.006$ $\mathrm{rad}$. This value is in good agreement with the general value given by \citet{Pushkarev2009}.

\begin{figure*}
\includegraphics[scale=0.37]{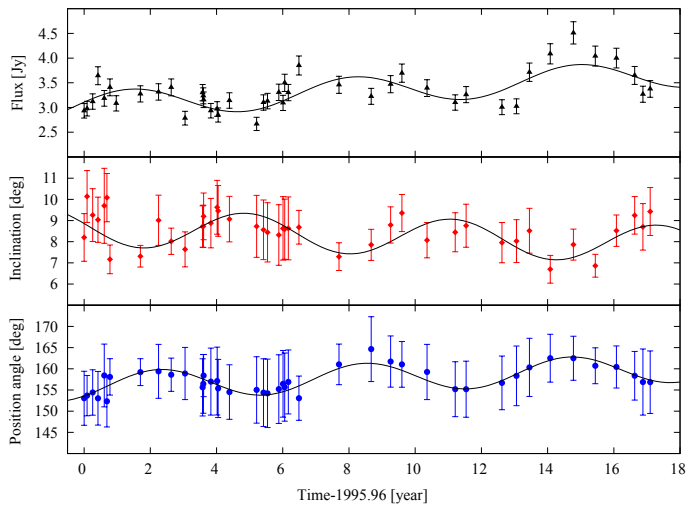}
\caption{Total flux density of the jet at $15$ GHz (top), inclination (middle) and
position angle (bottom) of the jet geometrical axis plotted against time. The flux
variability refers to the jet as a whole. The error bars represent the error
of the angles after the parametric fit given by Eq. (\ref{proj_geom1}) and Eq. (\ref{proj_geom2}) on the $xy$ positions of the inner jet. The curves are analytical fits to characterize the periodicities and linear trends in the time-series (parameters of the curves are given in Table \ref{fit_res}).}
\label{jetpers}
\end{figure*}

Next we fitted the measured component positions in all epochs of the MOJAVE data with Eq. (\ref{proj_geom1}) and Eq. (\ref{proj_geom2}). We used $a$ and $b$
parameters obtained from the $43$ GHz fit as fixed parameters. We allow the inclination and
position angle to vary from the starting values given in Table \ref{pars_mastercone}.
 The resulting time series of the inclination and position angle of the jet axis are given
in Table \ref{incli_ps_angles} and plotted in the lower two panels of Fig. \ref{jetpers}. The individual components can have smaller or larger inclination than the jet axis has, as the plasma flows around the jet axis, along the magnetic helix.
 
 \begin{table}
\begin{center}
\caption{The inclination and position angles of the jet axis obtained from the parametric fit on the 15 GHz data. The full table is available in electronic format online.}
\begin{tabular}{@{}c c c}
\hline
\hline
Epoch & $\lambda_0$ & $\iota_0$ \\
\hline
1995.96 & 153.05 $\pm$ 6.38 & 8.20 $\pm$ 1.13 \\ 
1996.05 & 153.67 $\pm$ 4.75 & 10.14 $\pm$ 1.23 \\ 
1996.22 & 154.41 $\pm$ 5.36 & 9.26 $\pm$ 1.25 \\ 
1996.38 & 153.04 $\pm$ 6.33 & 9.04 $\pm$ 1.07 \\ 
1996.57 & 158.42 $\pm$ 7.40 & 9.70 $\pm$ 1.77 \\ 
1996.74 & 152.34 $\pm$ 6.06 & 10.08 $\pm$ 1.14 \\ 
1996.82 & 158.09 $\pm$ 4.30 & 7.17 $\pm$ 0.68 \\
1996.93 & 155.77 $\pm$ 5.72 & 7.57 $\pm$ 1.00 \\
\hline
\hline
\end{tabular}
\label{incli_ps_angles}
\end{center}
\end{table} 
 
\begin{table*}
\begin{minipage}{140mm}
\begin{center}
\caption{The parameters of the analytical fits to the time-series of the total flux densities at $15$ GHz, inclination angles, and position angles (their mathematical form given in Eq. (\ref{iota_var}) and Eq. (\ref{an_fit})) are displayed. The total $\chi^2$ values are also indicated (the degrees of freedom are $N=39$ for all cases).}
\begin{tabular}{ccccc}
\hline
\hline
\multicolumn{5}{c}{$F$, $\chi^2=22.20$}\\
\hline
$A_0$[Jy]  & $A_1$[Jy]  & $A_2$ [Jy yr$^{-1}$] & T [yr] & $\phi$ [$\degr$] \\ 
3.03 $\pm$ 0.06 & 0.29 $\pm$ 0.06 & 0.04 $\pm$ 0.01  & 6.74 $\pm$ 0.24 & -15.44 $\pm$ 16.33\\
\hline
\multicolumn{5}{c}{$\iota_0$, $\chi^2=20.94$} \\
\hline
$A_0[\degr]$  & $A_1[\degr]$  & $A_2$ [$\degr$yr$^{-1}$] & T [yr] & $\phi$ [$\degr$]\\
8.67 $\pm$ 0.19 & 0.89 $\pm$ 0.17 & -0.05 $\pm$ 0.02 & 6.22 $\pm$ 0.19 & 11.95 $\pm$ 15.01\\
\hline
\multicolumn{5}{c}{$\lambda_0$, $\chi^2=5.82$}\\
\hline
$A_0[\degr]$  & $A_1[\degr]$  & $A_2$ [$\degr$yr$^{-1}$] & T [yr] & $\phi$ [$\degr$]\\
  155.90 $\pm$ 0.36 & 3.39 $\pm$ 0.33 & 0.24 $\pm$ 0.04 & 6.20$\pm$ 0.10 & 42.99$\pm$ 7.83\\
\hline
\hline
\label{fit_res}
\end{tabular}
\end{center}
\end{minipage}
\end{table*}

As it can be seen from Fig. \ref{jetpers}, the flux density, the inclination, and the position angle vary in a
periodic way around slowly changing average values. The orbital motion of the jet emitter BH explains the periodic behaviour. Our BBH model takes into account the monotonic change as well, via the spin-orbit precession. Therefore we fitted a harmonic function extended with a monotonic changing term to the data, that reflects a periodic variation with time while also allowing for linear drift in the average values:
\begin{flalign}
\label{flux_var}
&F(t)\!=\!A_0(F)\!+\!A_1(F)\sin \left(\frac{2\pi}{T(F)}t\!-\!\phi(F) \right)\!+\!A_2(F)t,\\
\label{iota_var}
&\iota _0(t)\!=\!A_0(\iota _0)\!+\!A_1(\iota _0)\sin \left(\frac{2\pi}{T(\iota _0)}t\!-\!\phi(\iota _0) \right)\!+\!A_2(\iota _0)t,\\
&\lambda _0(t)\!=\!A_0(\lambda _0)\!+\!A_1(\lambda _0)\sin \left(\frac{2\pi}{T(\lambda _0)}t\!-\!\phi(\lambda _0) \right)\!+\!A_2(\lambda _0)t.
\label{an_fit}
\end{flalign}
The numerical value of the parameters in the above functions were obtained by performing a non-linear least squares fitting on the time-variation of the corresponding jet-quantities using the Levenberg-Marquardt algorithm, such that the $\chi^2$ was minimized during the fitting. The coefficients of Eq. (\ref{flux_var}), Eq. (\ref{iota_var}), Eq. (\ref{an_fit}), and the corresponding $\chi^2$ values are given in Table \ref{fit_res}.

In Section \ref{SMBH binary in the central part of the S5 1928+738} we will use the inclination parametrization given by Eq. (\ref{iota_var}) to deduce the binary parameters. For this reason a hypothesis test was carried out to check whether the function given in Eq. (\ref{iota_var}), predicted by our BBH model, is consistent with the inclination variation of the jet axis. The null-hypothesis states that the inclination-variations of the jet axis at different epochs are drawn from a random distribution. The alternative hypothesis states that the variations are drawn from a distribution described by Eq. (\ref{iota_var}).
We checked the probability $P(\chi^2<={\chi^2}_\mathrm{cv})$, where ${\chi^2}_{cv}$ is 20.94 for $\iota_0$, by performing a non-parametric Monte-Carlo test. We produced $10000$ new time-series by randomly mixing the original $44$ inclination data points and then fitted these with Eq. (\ref{iota_var}). The test gave $\chi^2<\chi^2_{\mathrm{cv}}$ in $24$ cases implying $P(\chi^2<\chi^2_\mathrm{cv})\approx0.0024$. Thus we rejected the null-hypothesis, and accepted that the fit mirrors real variations. We can conclude that our BBH model (periodic variation due to orbital motion, and monotonic change due to spin-orbit precession) is able to explain the inclination angle variation of the jet axis.
 
 The variation in the inclination angle of the jet axis permits that $\iota_0$ changes between approximately $7^\circ$ and $9^\circ$, additionally the non-zero jet opening angle of $\psi^\mathrm{int}$ allows the inclination angle of the components to be between $6^\circ$ and $10^\circ$. The minimal and maximal apparent velocities are $3.4c$ and $6.6c$ in the first $2$ mas part of the jet, respectively. Assuming a constant Lorentz-factor of $\Gamma=8.1$ for the whole jet, the observed apparent velocities cannot be reproduced.
 Changes in the intrinsic jet velocity are often observed in extragalactic jets \citep[e.g.][and references therein]{Asada2014}. 
In the present case, the intrinsic jet velocity should be between $\approx0.965c$ and $\approx0.990c$ allowing a variable $\beta_\mathrm{jet}$.

\subsection{Radio flux density variability}
\label{Consistency check through radio flux density variability}

The total flux density of the jet is plotted against time on the upper panel in Fig. \ref{jetpers}. We fitted the data by a function similar to those given for the inclination and position angles. The resulting coefficients are given in the first line of Table \ref{fit_res}. The total flux density variability can be described in a way that it is brighter (fainter) when the average inclination angle of the jet axis is smaller (larger). This points towards variable Doppler boosting as the reason behind the observed radio flux variability.

In the case of S5 1928+738, the total flux density variation at $15$ GHz is dominated by components Cg and CS. The flux variability of the core region indicates a six year periodicity, similar to the period of the inclination variation of the jet axis. The monotonic decreasing value of the average inclination angle suggests that the jet axis moves closer and closer to the line of sight. This slow reorientation of the inner jet is consistent with the observed long-term brightening of the jet.

On the other hand, the sharp peaks in the flux density of the core region (at epochs 2002.45 yr and 2010.04 yr) could also be attributed to component ejections, following the shock-in-jet model of \citet{Marscher1985} and \citet{Bjornsson2000}. The shock-shock interaction, when a newly ejected component travels through the core region of the jet may cause the observed brightening. The components having closest ejection epochs to the peaks are C12 ($t_\mathrm{ej,C12}=(2000.74\pm0.19)\mathrm{yr}$) and C16 ($t_\mathrm{ej,C16}=(2008.83\pm0.20)\mathrm{yr}$). If we assume that the components maintained their observed speed ($\mu_{C12}=0.180\pm0.009$ mas yr$^{-1}$ and $\mu_\mathrm{C16}=0.180\pm0.013$ mas yr$^{-1}$) after their ejections, they arrived at the position of Cg approximately at the time of the flux density increase. In this scenario Cg is actually a standing shock zone in the jet, that brightens as a powerful moving shock (components C12 and C16) moves through it.

It is plausible that the radio flux variability of the core region of S5 1928+738 at $15$ GHz has a composite origin. The component ejections can explain the periodic component of the flux variability. On the other hand, its linear component can be attributed to the manifestation of the spin precession of the jet emitter black hole through the Doppler boosting of the emitted light.

\section{SMBH binary in the central part of the S5 1928+738}
\label{SMBH binary in the central part of the S5 1928+738}

In this section we connect the presented jet behaviour to a dynamical
model of merging SMBH binary. Following \citet{Roos1993} we assume that the periodic change of the polar angle of the jet axis is generated by a hidden
SMBH binary at the base of the jet. Then we identify the linear
trend in the evolution of the average angles by the slow reorientation of the spin of the jet
emitter BH due to the spin-orbit precession. Based on the above assumptions
we calculate the binary parameters. Note, that in this section we do not consider the helix described in the geometrical
model, rather we consider its axis that itself spirals due to the orbital motion. This secondary spiral structure is revealed by the
periodically changing inclination and position angles of the jet axis (see Fig. \ref{jetpers}).

\subsection{Earlier estimation of total mass, binary separation and orbital period}
\label{totalmass}

Once the immediate environment of the central mass becomes sufficiently active to radiate like a QSO, its
luminosity $L$ approaches the Eddington luminosity $L_\mathrm{E}$ \citep[for a review see e.g.][]{Schneider2006}.
\citet{Roos1993} adopted a total mass of $m_{\mathrm{I}}\approx 10^8 M_{\odot}$ for the SMBH binary hosted in S5 1928+738, based on the Eddington limit and assuming a bolometric luminosity of $8.15 \cdot 10^{45} \mathrm{erg s}^{-1}$. Further mass estimates have been published by other authors, constraining the total mass of the SMBH binary at around $10^8 M_{\odot}$ \citep[][]{Kelly2007,Liu2006}. An order of magnitude higher SMBH mass was estimated by \citet{Cao2002}. With the large uncertainties and intrinsic scatter of SMBH mass estimates from different methods noted, here we adopt $m\approx8.13 \times 10^{8} M_{\odot}$ derived by \citet{Woo2002} from the black hole mass-AGN continuum luminosity scaling relation, which in turn is based on the reverberation mapping technique.

\citet{Hummel1992} fitted a moving sine-wave on the component positions of the mas-scale jet of S5 1928+738 obtaining a wavelength of $\sim1.06$ mas, phase shift of $\sim0.28$ mas yr$^{-1}$ and amplitude of $\sim0.09$ mas. \citet{Roos1993} estimated the orbital period of the binary as the wavelength of this sine-wave divided by its phase-shift, yielding an orbital period of $\sim2.9$ yr, measured in the rest frame of the source.
By considering the total mass $m\sim 10^8 M_{\odot}$ and the orbital period $\sim2.9$ yr, \citet{Roos1993} calculated the binary separation of $r\approx0.003$ pc based on Kepler's third law. Compared to these, we adopt a refined model, paying attention to details related to the spin.

\subsection{Binary parameters}
\label{Binary parameters}

We consider two SMBHs orbiting each other with total mass $m=m_1+m_2$ and mass ratio $\nu=m_2/m_1 < 1$.
The equation of motion is \citep{Kidder1995}: 
\begin{flalign}
\frac{d^2\mathbf{r}}{dt^2}\!=\!-\frac{m\mathbf{r}}{r^3} \left(1+\mathcal{O}(\varepsilon)+\mathcal{O}(\varepsilon^{1.5})+\mathcal{O}(\varepsilon ^{2})+\mathcal{O}(\varepsilon ^{2.5})+...\right),
\label{eqmotion}
\end{flalign}
where \textbf{r} is the binary separation, $r=\vert\mathbf{r}\vert$, $\varepsilon=Gmc^{-2}r^{-1}$ is the post-Newtonian (PN) parameter and $\mathcal{O}(\varepsilon ^{n})$ represents the \textit{n}th PN order. The radial evolution of the binary is characterized to Newtonian order by the total mass $m$, the orbital period $T$ and the separation $r$, only two of them being independent. To 1PN accuracy the mass ratio enters as a fourth parameter.

If the jet periodicities observed in S5 1928+738 are consequences of the orbital motion of the binary, then the separation can be calculated to Newtonian order as: 
\begin{flalign}
r\!=\!0.29 \times (Gm)^{1/3} T^{2/3},
\label{kepler3}
\end{flalign}
where $T$ is provided from the periodic component in the variation of the inclination $T(\iota_0)$, so that $T=T(\iota_0)(1+z)^{-1}=4.78\pm0.14$ yr in the rest frame of the source. By employing the total mass estimate of \citet{Woo2002} $m=8.13\times 10^{8} M_{\odot}$, Eq. (\ref{kepler3}) implies the separation $r=0.0128\pm0.0003$ pc and consequently the PN parameter $\varepsilon\approx0.003$.

After the gravitational radiation
becomes the dominant dissipative effect over the dynamical friction, the evolution of the compact binary passes through three phases, the inspiral, the plunge, and the ring-down. The $\varepsilon \approx 0.003$ value of the PN parameter found in S5 1928+738 indicates that the SMBH binary is in the inspiral phase, typically encompassing the range $0.001<\varepsilon<0.1$ \citep{Gergely2009,Levin2011}.
 
 Up to 2PN order the merger dynamics are conservative, the constants of motion being the total energy E and the angular momentum vector $\mathbf{J}=\mathbf{S_1}+\mathbf{S_2}+\mathbf{L}$, where $\mathbf{L}$ is the orbital angular momentum, $\mathbf{S_1}$ is the spin of the dominant SMBH and $\mathbf{S_2}$ is the spin of the secondary SMBH. The SMBH spins obey precessional motion \citep{Barker1975,Barker1979}:
\begin{flalign}
\mathbf{\dot{S}_i}\!=\!\Omega_i \times \mathbf{S_i},
\end{flalign}
where index $i$ refers to the first and second member of the binary, the angular velocity $\Omega_i$ of the \textit{i}th spin $\mathbf{S_i}$ contains up to 2PN order spin-orbit (1.5PN), spin-spin (2PN), and quadrupole momentum contributions (2PN). For the typical mass ratio only the dominant spin is important \citep{Gergely2009}, and the spin of the secondary black hole $\mathbf{S_2}$ can be neglected. The 2PN effects are also neglected. In the one-spin case the dominant spin $\mathbf{S}$ precesses around $\mathbf{J}$ with the angular velocity \citep{Gergely2009}:
\begin{flalign}
\Omega_{\mathrm{p}}\!=\!\frac{2c^3}{Gm} \varepsilon^{5/2} \frac{\nu}{(1+\nu)^2}.
\label{omegaso}
\end{flalign}
 Reorientation of the spin of the dominant SMBH should consequently cause variations in the jet direction. We identify the linear change in the direction of jet axis of S5 1928+738 with the spin precession of the dominant SMBH. 

Once the binary evolution has progressed into the inspiral stage, the black
holes approach further and coalescence is expected within the gravitational
time-scale of the system \citep{Gergely2009}:
\begin{flalign}
T_{\mathrm{merger}}\!=\!\frac{5Gm}{32c^{3}}\varepsilon ^{-4} \frac{(1+\nu)^2}{\nu},
\label{tmerger}
\end{flalign}
due to gravitational wave emission. The semi-major axis and eccentricity of the elliptical orbits both decrease due to the emitted gravitational radiation, the latter occurring faster, hence circularizing the orbit \citep{Peters1964}, what we assume hereafter.
 
\begin{figure}
\begin{center}
\includegraphics[scale=0.45]{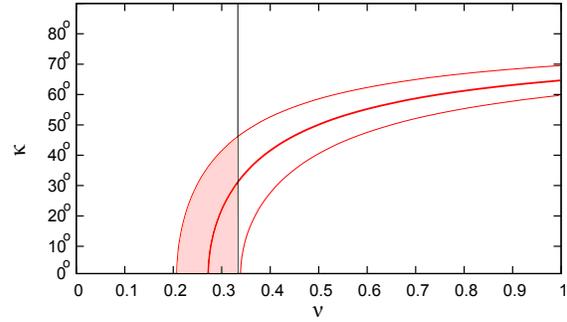}
\end{center}
\caption{The thick solid line represents the function $\kappa(\nu)$ given by Eq. (\ref{kappa_nu}),
with the upper and lower continuous lines representing
the respective errors. The vertical solid line sets the upper
limit of the typical mass ratio at $\nu=1/3$ (a higher mass ratio would imply comparable spins, hence two jets, contradicting the observations). The shaded area indicates the parameter range allowed by observations of the jet. Although the available data on the function $\kappa(\nu)$ does not constrain the precession angle from below, a
vanishing $\kappa$ is not allowed, as it would switch off the spin-orbit
precession.}
\label{mr}
\end{figure}

Next we determine the mass ratio interval compatible with observations in our model.
The orbital motion of the jet emitter black hole makes the jet geometrical axis wiggle about the spin with a half-opening angle $\zeta$. This angle is the amplitude of the periodic component in $\iota_0(t)$ determined as $\zeta=A_1(\iota_0)(1+z)^{-1}=0.68\degr \pm 0.13 \degr$. It can be also expressed as:
\begin{flalign}
\zeta\!=\! \arcsin \frac{v_{1}\cos \kappa }{v_{\mathrm{jet}}},
\label{psiint}
\end{flalign}
where $v_{1}\cos \kappa$ is the component of the orbital velocity of the jet emitter BH perpendicular to the spin direction, $\kappa$ represents the angle between the spin of the jet emitter black hole and the orbital angular momentum. Substituting the orbital velocity of the jet emitter black hole $\beta_1=v_1 c^{-1}=\varepsilon^{1/2} \nu (1+\nu)^{-1}$, and the intrinsic jet velocity $\beta_{\mathrm{jet}}$ (both given in units of $c$) into the above equation, the angle $\kappa$ can be expressed as:
\begin{flalign}
\kappa=\arccos \frac{\beta_{\mathrm{jet}} \sin \zeta}{\varepsilon^{1/2}} \frac{1+\nu}{\nu}.
\label{kappa_nu}
\end{flalign}
Substituting the known parameters into the above equation, the angle $\kappa$ becomes a function of the mass ratio. 
We plot $\kappa(\nu)$ in Fig. \ref{mr}. The parameter space is restricted below.

 The SMBHs in the binary co-evolved in similar astrophysical environments, therefore it is plausible to
assume that they have similar dimensionless spin parameter, hence $S_2/S_1\approx\nu^{2}$. Values of $\nu>1/3$ and consequently $S_2/S_1>0.1$ would allow for two pronounced jets in the system (morphologically like the X-shaped radio sources). Nevertheless there is no evidence of a double jet base in the jet, which would generate different families of component motions, indicating spins of comparable magnitude. Therefore we restrict the mass ratio to $\nu<1/3$.

When the spin of a Kerr black hole is non-parallel to the orbital angular momentum, the spin-orbit interaction leads to orbital plane and spin precessions. The slow reorientation and brightening of the inner jet strongly supports such a precession, hence parallel dominant spin and orbital angular momentum configurations with $\kappa=0\degr$ are ruled out.

As it can be seen from Fig. \ref{mr}., taking into account the
$1\sigma$ error of the inclination amplitude $\mathrm{d}\zeta=0.13 \degr$ and of the PN parameter $\mathrm{d}\varepsilon=7\times10^{-5}$, the above arguments constrain the mass ratio to the range $\nu \in \left[ 0.21:1/3\right] $. We note, that the expected small change in $\beta_\mathrm{jet}$ along the jet ($\Delta\approx0.03c$) would just slightly modify the limit on $\nu$ (see Eq. (\ref{psiint})). Then the spin-orbit precession period $T_{\mathrm{SO}}=4852 \pm 646$ yr is estimated via Eq. (\ref{omegaso}) and the merger time $T_{\mathrm{merger}}=(1.44 \pm 0.19)\times 10^6$ yr is estimated via Eq. (\ref{tmerger}). The binary parameters are summarized in Table \ref{binarypars}.

\begin{table}
\caption{Binary parameters. $^\star$:independent result by \citet{Woo2002}.}
\begin{tabular}{@{}l l }
\hline
\hline
Total mass, m$^\star$ [$M_{\odot}$]  & ${8.13  \times 10^{8}}$\\
Orbital period, T [yr] & $4.78\pm0.14$\\
Binary separation, $r$ [pc] & $0.0128 \pm 0.0003$ \\
PN parameter, $\varepsilon$ & $\approx 0.003$ \\
Mass ratio, $\nu$ & $\left[0.21:1/3\right]$\\ 
Spin-orbit precession period, $T_{\mathrm{SO}}$ [yr] & $4852 \pm 646$ \\
Gravitational lifetime, $T_{\mathrm{merger}}$ [yr] & $(1.44 \pm 0.19)\times 10^6$\\
\hline
\hline
\end{tabular}
\label{binarypars}
\end{table}

\section{Discussion and Conclusions}

In this paper we investigated the radio jet of S5 1928+738 based on calibrated data of the MOJAVE survey with focus on the perturbed jet ejection. Our analysis basically confirmed the model of \citet{Roos1993} which in turn was based on the jet analysis by \citet{Hummel1992} of a hidden SMBH binary at the jet base, causing the observed wiggling of the jet. The improvement upon this model, advanced in the present paper is the inclusion of the spin of the jet-producing black hole.

We developed a detailed geometrical model to describe the inner $2$ mas of the jet by using the
$43$ GHz VLBA-map of \citet{Lister2000}. From that we obtained the inclination and position angle of the jet
axis and the intrinsic half-opening angle of the conical helix. Then we fitted the geometric model to the $15$ GHz VLBA data of almost twenty years. Three features of the jet have been revealed; (i) a conical helix
shape, (ii) a periodical change in the direction of its symmetry axis, (iii)
a slow additional reorientation in the average direction of the jet. Our spinning binary black hole model naturally explained the simultaneous presence of properties (ii) and (iii), and (i) was attributed to the presence of helical kink instabilities.

Jet precession (iii) can be induced through the Bardeen-Petterson interaction between a viscous accretion disk and a spinning black hole \citep{BardeenPetterson1975}, if the disk is misaligned compared to the equatorial plane of the black hole. 
However such a scenario cannot explain both (ii) and (iii). In an alternative scenario, the influence of the immediate environment could cause the wiggling of the jet. However, to explain the observed periodicities would require properly fine-tuned structures. So far there is no evidence for such periodic distribution of dense material close to the jet.

 We adopted the total mass of the binary $m=8.13 \times 10^{8}M_{\odot }$  determined by \citet{Woo2002} from the black hole mass-AGN continuum luminosity scaling relation (this value is different by a factor of $8.13$ from the one adopted
by \citealt{Roos1993}). The helical jet model with periodic jet axis and VLBI data implied the orbital period $T=4.78\pm 0.14$ yr (this is a factor of $1.59$  larger than the value given by \citealt{Roos1993}). With these values we calculated the binary separation as $r=0.0128\pm 0.0003$ pc. These parameters imply that the SMBH binary is in the inspiral phase, but far from coalescence, with post-Newtonian parameter $\varepsilon \approx 0.003$.

Long-term monitoring of the radio jet allowed us to identify the linear trend in the evolution of the inclination and position angles of the jet axis, interpreted as arising from the spin-orbit precession of the jet emitter SMBH.
The mass ratio most likely falls into the range $\nu \in \left[ 0.21:1/3 \right] $. The spin-orbit precession period was identified as $T_{\mathrm{SO}}=4852 \pm 646$ yr and the
gravitational lifetime emerged as $T_{\mathrm{merger}}=(1.44 \pm 0.19)\times 10^6$ yr.

Although we cannot rule out that other models could explain the observed jet structure, we showed that the VLBI data of S5 1928+738, extending over almost twenty years is consistent with the model of a spinning binary black hole lying at the jet base, where the larger black hole has a spin detectable through its spin-orbit precession. Measurements of slow increase in the average flux density of the jet in the newest three epochs (2013.34, 2013.58 and 2013.96) further support the model, as such an increase is predicted by spin-orbit precession. Our study thus provides indications, for the first time from VLBI jet kinematics, for the spinning nature of the jet-emitting black hole.

 As the \-MOJAVE survey is still ongoing, further data on S5 1928+738 may better constrain the parameters of the model. With a significantly increased amount of data in principle it would be possible to monitor higher order post-Newtonian effects too. Beyond spin induced precession, such an analysis could also reveal the magnitude of the spin, unavailable at the accuracy of the present analysis.
 
\section*{Acknowledgements}
We thank the referee for helpful comments and suggestions. EK was partially supported by COST Action MP0905 "Black Holes in a Violent Universe". EK acknowledges financial support from the International Space Science Institute. During this research, K\'{E}G and L\'{A}G were supported by the European Union and the State of Hungary, co-financed by the European Social Fund in the framework of T\'{A}MOP-4.2.4.A/2-11/1-2012-0001 "National Excellence Program". In the early stages of this research K\'{E}G was supported by the Hungarian OTKA grant K 104539 and L\'{A}G by EU grant T\'{A}MOP-4.2.2.A-11/1/KONV-2012-0060 and the Japan Society for the Promotion of Science. MK was
supported by the National Research Foundation of Korea (NRF) grant, No.
2008-0060544, funded by the Korea government (MSIP). This research has made use of data from the MOJAVE database that is maintained by the MOJAVE team. The Very Long Baseline Array (VLBA) is an instrument of the National Radio Astronomy Observatory (NRAO). The National Radio Astronomy Observatory is a facility of the National Science Foundation operated under cooperative agreement by Associated Universities, Inc.

 \newpage

\onecolumn
\begin{table*}
\begin{minipage}{140mm}
\begin{center}
\caption{Summary of the 15 GHz image parameters. (1) epoch of the VLBA observation, (2) VLBA experiment code, (3)--(4) FWHM minor and major axis of the restoringbeam , respectively, (5) position angle of the major axis of the restoring beam measured from North through East, (6) rms noise of the image, (7) reduced $\chi{^2}$ of the \textsc{Difmap} model-fit, (8) number of the components in the model.}
\centering
\begin{tabular}{cccccccc}
\hline
\hline
	&	VLBA  & $\mathrm{B}_{\mathrm{min}}$ & $\mathrm{B}_{\mathrm{maj}}$& $\mathrm{B}_{\mathrm{PA}}$ & rms  & Red. & Comp.\\
Epoch & Code & [mas] & [mas] & [$^{\circ}$] & mJy bm$^{-1}$ & $\chi{^2}$& Number \\
(1) & (2) &  (3) &  (4) &  (5) &  (6) & (7) & (8)\\
\hline
1994 Aug 31 & BZ004 & 0.67 & 0.76 & -67 & 0.4 & 1.21 & 7\\
1995 Dec 15 & BK37A & 0.53 & 0.76 & 12 & 0.7 & 0.39 & 11\\
1996 Jan 19 & BR034 & 0.44 & 1.00 & -9 & 0.6 & 1.16 & 11\\
1996 Mar 22 & BR034B & 0.46 & 1.08 & -16 & 0.4 & 1.01 & 11\\
1996 May 16 & BK037B & 0.52 & 0.83 & -39 & 0.3 & 1.39 & 11\\
1996 Jul 27 & BR034D & 0.48 & 0.97 & -16 & 0.4 & 2.55 & 10\\
1996 Sep 27 & BR034E & 0.47 & 0.91 & -43 & 0.4 & 1.19 & 8\\
1996 Oct 27 & BK037D & 0.42 & 0.65 & 22 & 0.5 & 0.85 & 11\\
1996 Dec 06 & BR034F & 0.48 & 1.00 & -25 & 0.5 & 1.47 & 8\\
1997 Aug 28 & BK052A & 0.54 & 0.76 & -1 & 0.5 & 0.47 & 12\\
1998 Mar 19 & BK052C & 0.54 & 0.85 & 15 & 0.5 & 0.44 & 11\\
1998 Aug 03 & BT039 & 0.54 & 0.71 & -8 & 0.3 & 1.19 & 12\\
1999 Jan 02 & BG077D & 0.56 & 0.64 & -13 & 0.5 & 0.44 & 12\\
1999 Jul 17 & BP053 & 0.60 & 0.70 & -49 & 0.2 & 1.27 & 12\\
1999 Jul 24 & BA037 & 0.64 & 0.69 & 56 & 0.3 & 1.23 & 12\\
1999 Jul 26 & BM114B & 0.55 & 1.29 & 14 & 0.3 & 7.56 & 12\\
1999 Oct 16 & BA037B & 0.56 & 0.64 & 49 & 0.3 & 1.26 & 12\\
1999 Dec 23 & BA037C & 0.55 & 0.64 & 59 & 0.3 & 1.36 & 12\\
2000 Jan 02 & BP053B & 0.55 & 0.74 & -23 & 0.1 & 1.29 & 12\\
2000 May 08 & BP053C & 0.58 & 0.71 & -13 & 0.1 & 1.27 & 12\\
2001 Mar 04 & BK068E & 0.53 & 0.80 & -6 & 0.3 & 2.58 & 12\\
2001 May 17 & BT056 & 0.54 & 0.74 & -12 & 0.4 & 1.47 & 10\\
2001 Jun 30 & BA051A & 0.54 & 0.76 & 16 & 0.3 & 1.25 & 11\\
2001 Nov 02 & BA051B & 0.53 & 0.75 & 18 & 0.3 & 1.30 & 11\\
2001 Dec 22 & BR077G & 0.53 & 0.80 & -13 & 0.2 & 2.45 & 10\\
2002 Jan 07 & BA051C & 0.57 & 0.81 & 24 & 0.3 & 1.25 & 9\\
2002 Feb 18 & BR077K & 0.55 & 0.83 & 5 & 0.2 & 3.14 & 9\\
2002 Jun 15 & BL111B & 0.52 & 0.66 & -56 & 0.3 & 1.15 & 10\\
2003 Aug 28 & BL111J & 0.61 & 0.70 & 40 & 0.3 & 1.19 & 11\\
2004 Aug 19 & BM209B & 0.56 & 1.11 & -75 & 0.3 & 1.50 & 10\\
2005 Mar 23 & BL123D & 0.59 & 0.65 & -74 & 0.3 & 1.24 & 10\\
2005 Jul 24 & BL123J & 0.57 & 0.69 & -35 & 0.2 & 1.19 & 10\\
2006 Apr 28 & BL137D & 0.59 & 0.72 & -58 & 0.4 & 1.18 & 10\\
2007 Mar 02 & BL137O & 0.65 & 0.72 & -52 & 0.3 & 1.13 & 10\\
2007 Jul 03 & BL149AC & 0.64 & 0.67 & 39 & 0.2 & 1.26 & 10\\
2008 Jul 30 & BL149AL & 0.62 & 0.65 & 34 & 0.2 & 1.44 & 10\\
2009 Jan 07 & BL149BG & 0.64 & 0.71 & -85 & 0.2 & 1.21 & 10\\
2009 May 28 & BL149BL & 0.62 & 0.66 & -69 & 0.2 & 1.23 & 10\\
2010 Jan 16 & BL149CG & 0.67 & 0.77 & 79 & 0.2 & 1.13 & 10\\
2010 Sep 29 & BL149CR & 0.60 & 0.63 & 21 & 0.2 & 1.29 & 10\\
2011 May 26 & BL149DI & 0.61 & 0.66 & -10 & 0.2 & 1.17 & 11\\
2012 Jan 14 & BL178AF & 0.68 & 0.75 & 42 & 0.2 & 1.06 & 11\\
2012 Aug 03 & BL178AN & 0.65 & 0.69 & 53 & 0.1 & 1.93 & 11\\
2012 Nov 02 & BL178AR & 0.63 & 0.68 & 73 & 0.2 & 1.19 & 12\\
2013 Jan 21 & BL178AY & 0.67 & 0.71 & -86 & 0.2 & 1.13 & 12\\
\hline
\hline
\end{tabular}
\end{center}
\end{minipage}
\end{table*}

\clearpage
\newpage

\begin{center}
\begin{longtable}{lccccc}
\caption{Circular Gaussian model-fit results for S5 1928+738. (1) epoch of observation, (2) flux density, (3)--(4) position of the component center with respect to the core, (5) FWHM, (6) jet-component identification.}\\

\hline
Epoch [yr] & Flux density [Jy] & x [mas] & y [mas] & d [mas]  & CO\\ \hline 
\endfirsthead

\multicolumn{6}{c}%
{{\tablename\ \thetable{} -- \textit{Continued from previous page}}} \\ \hline
\endhead
1994.67 & 0.606 $\pm$ 0.044 & 0.000 $\pm$ 0.040 & 0.000 $\pm$ 0.037 & 0.145 $\pm$ 0.003 & CS\\
 & 0.462 $\pm$ 0.045 & 0.227 $\pm$ 0.040 & -0.311 $\pm$ 0.037 & 0.144 $\pm$ 0.005 & Cg\\
 & 0.407 $\pm$ 0.042 & 0.561 $\pm$ 0.048 & -1.238 $\pm$ 0.045 & 0.301 $\pm$ 0.007 & C7\\
 & 0.408 $\pm$ 0.042 & 0.258 $\pm$ 0.051 & -1.613 $\pm$ 0.049 & 0.350 $\pm$ 0.007 & C6\\
 & 0.078 $\pm$ 0.020 & 1.153 $\pm$ 0.087 & -3.282 $\pm$ 0.086 & 0.789 $\pm$ 0.064 & C3\\
 & 0.038 $\pm$ 0.022 & 2.203 $\pm$ 0.206 & -7.417 $\pm$ 0.206 & 2.029 $\pm$ 0.580 & C2\\
 & 0.035 $\pm$ 0.012 & 2.675 $\pm$ 0.142 & -10.212 $\pm$ 0.141 & 1.366 $\pm$ 0.175 & C1\\
\hline
1995.96 & 0.631 $\pm$ 0.039 & 0.000 $\pm$ 0.027 & 0.000 $\pm$ 0.037 & 0.020 $\pm$ 0.002 & CS\\
 & 0.402 $\pm$ 0.031 & 0.086 $\pm$ 0.027 & -0.213 $\pm$ 0.037 & 0.038 $\pm$ 0.003 & Cg\\
 & 0.781 $\pm$ 0.046 & 0.348 $\pm$ 0.028 & -0.609 $\pm$ 0.038 & 0.063 $\pm$ 0.002 & C8\\
 & 0.151 $\pm$ 0.020 & 0.359 $\pm$ 0.031 & -0.848 $\pm$ 0.040 & 0.146 $\pm$ 0.008 & B1\\
 & 0.416 $\pm$ 0.043 & 0.715 $\pm$ 0.036 & -1.616 $\pm$ 0.044 & 0.236 $\pm$ 0.005 & C7\\
 & 0.233 $\pm$ 0.032 & 0.246 $\pm$ 0.043 & -1.928 $\pm$ 0.050 & 0.330 $\pm$ 0.010 & C6\\
 & 0.224 $\pm$ 0.032 & 0.366 $\pm$ 0.042 & -2.195 $\pm$ 0.049 & 0.324 $\pm$ 0.011 & C5\\
 & 0.049 $\pm$ 0.012 & 1.271 $\pm$ 0.063 & -3.345 $\pm$ 0.067 & 0.564 $\pm$ 0.043 & C4\\
 & 0.025 $\pm$ 0.009 & 1.314 $\pm$ 0.071 & -4.222 $\pm$ 0.075 & 0.654 $\pm$ 0.085 & C3\\
 & 0.020 $\pm$ 0.008 & 2.768 $\pm$ 0.082 & -8.220 $\pm$ 0.086 & 0.779 $\pm$ 0.130 & C2\\
 & 0.052 $\pm$ 0.022 & 2.932 $\pm$ 0.215 & -10.709 $\pm$ 0.216 & 2.128 $\pm$ 0.352 & C1\\
\hline
1996.05 & 0.160 $\pm$ 0.035 & 0.000 $\pm$ 0.023 & 0.000 $\pm$ 0.048 & 0.067 $\pm$ 0.033 & CS\\
 & 0.971 $\pm$ 0.086 & 0.169 $\pm$ 0.023 & -0.335 $\pm$ 0.048 & 0.046 $\pm$ 0.006 & Cg\\
 & 0.741 $\pm$ 0.074 & 0.457 $\pm$ 0.023 & -0.832 $\pm$ 0.048 & 0.043 $\pm$ 0.007 & C8\\
 & 0.225 $\pm$ 0.041 & 0.518 $\pm$ 0.031 & -1.027 $\pm$ 0.052 & 0.209 $\pm$ 0.025 & B2\\
 & 0.417 $\pm$ 0.045 & 0.820 $\pm$ 0.037 & -1.866 $\pm$ 0.056 & 0.294 $\pm$ 0.009 & C7\\
 & 0.289 $\pm$ 0.037 & 0.357 $\pm$ 0.038 & -2.176 $\pm$ 0.057 & 0.306 $\pm$ 0.013 & C6\\
 & 0.129 $\pm$ 0.025 & 0.551 $\pm$ 0.033 & -2.600 $\pm$ 0.054 & 0.248 $\pm$ 0.028 & C5\\
 & 0.024 $\pm$ 0.010 & 1.463 $\pm$ 0.041 & -3.513 $\pm$ 0.059 & 0.342 $\pm$ 0.128 & C4\\
 & 0.050 $\pm$ 0.014 & 1.293 $\pm$ 0.085 & -4.291 $\pm$ 0.094 & 0.815 $\pm$ 0.087 & C3\\
 & 0.029 $\pm$ 0.020 & 1.940 $\pm$ 0.178 & -8.667 $\pm$ 0.183 & 1.769 $\pm$ 0.694 & C2\\
 & 0.051 $\pm$ 0.019 & 3.369 $\pm$ 0.182 & -10.768 $\pm$ 0.187 & 1.811 $\pm$ 0.265 & C1\\
\hline
1996.22 & 0.362 $\pm$ 0.058 & 0.000 $\pm$ 0.025 & 0.000 $\pm$ 0.047 & 0.073 $\pm$ 0.021 & CS\\
 & 0.810 $\pm$ 0.086 & 0.115 $\pm$ 0.025 & -0.254 $\pm$ 0.047 & 0.060 $\pm$ 0.009 & Cg\\
 & 0.713 $\pm$ 0.057 & 0.384 $\pm$ 0.026 & -0.698 $\pm$ 0.048 & 0.097 $\pm$ 0.005 & C8\\
 & 0.293 $\pm$ 0.037 & 0.465 $\pm$ 0.030 & -0.964 $\pm$ 0.050 & 0.178 $\pm$ 0.013 & B3\\
 & 0.387 $\pm$ 0.038 & 0.773 $\pm$ 0.037 & -1.816 $\pm$ 0.055 & 0.284 $\pm$ 0.008 & C7\\
 & 0.249 $\pm$ 0.034 & 0.282 $\pm$ 0.039 & -2.048 $\pm$ 0.056 & 0.304 $\pm$ 0.016 & C6\\
 & 0.182 $\pm$ 0.029 & 0.451 $\pm$ 0.038 & -2.480 $\pm$ 0.055 & 0.290 $\pm$ 0.022 & C5\\
 & 0.053 $\pm$ 0.017 & 1.353 $\pm$ 0.073 & -3.471 $\pm$ 0.083 & 0.687 $\pm$ 0.101 & C4\\
 & 0.026 $\pm$ 0.015 & 1.454 $\pm$ 0.069 & -4.622 $\pm$ 0.080 & 0.644 $\pm$ 0.292 & C3\\
 & 0.059 $\pm$ 0.040 & 2.515 $\pm$ 0.188 & -9.207 $\pm$ 0.192 & 1.861 $\pm$ 0.722 & C2\\
 & 0.031 $\pm$ 0.026 & 2.920 $\pm$ 0.117 & -11.484 $\pm$ 0.124 & 1.144 $\pm$ 0.740 & C1\\
\hline
1996.38 & 0.565 $\pm$ 0.044 & 0.000 $\pm$ 0.030 & 0.000 $\pm$ 0.033 & 0.050 $\pm$ 0.003 & CS\\
 & 0.872 $\pm$ 0.055 & 0.091 $\pm$ 0.031 & -0.192 $\pm$ 0.033 & 0.060 $\pm$ 0.002 & Cg\\
 & 0.822 $\pm$ 0.053 & 0.365 $\pm$ 0.031 & -0.646 $\pm$ 0.034 & 0.093 $\pm$ 0.002 & C8\\
 & 0.268 $\pm$ 0.030 & 0.406 $\pm$ 0.034 & -0.874 $\pm$ 0.036 & 0.153 $\pm$ 0.007 & B4\\
 & 0.457 $\pm$ 0.042 & 0.740 $\pm$ 0.043 & -1.734 $\pm$ 0.045 & 0.309 $\pm$ 0.005 & C7\\
 & 0.264 $\pm$ 0.032 & 0.247 $\pm$ 0.042 & -2.058 $\pm$ 0.044 & 0.299 $\pm$ 0.009 & C6\\
 & 0.205 $\pm$ 0.028 & 0.449 $\pm$ 0.049 & -2.380 $\pm$ 0.051 & 0.385 $\pm$ 0.012 & C5\\
 & 0.069 $\pm$ 0.015 & 1.278 $\pm$ 0.077 & -3.505 $\pm$ 0.078 & 0.710 $\pm$ 0.043 & C4\\
 & 0.028 $\pm$ 0.010 & 1.434 $\pm$ 0.077 & -4.665 $\pm$ 0.078 & 0.710 $\pm$ 0.106 & C3\\
 & 0.067 $\pm$ 0.022 & 2.643 $\pm$ 0.179 & -8.760 $\pm$ 0.179 & 1.760 $\pm$ 0.184 & C2\\
 & 0.051 $\pm$ 0.017 & 2.953 $\pm$ 0.159 & -11.306 $\pm$ 0.160 & 1.566 $\pm$ 0.170 & C1\\
\hline
1996.57 & 1.101 $\pm$ 0.086 & 0.000 $\pm$ 0.025 & 0.000 $\pm$ 0.044 & 0.047 $\pm$ 0.004 & CS\\
 & 0.269 $\pm$ 0.043 & 0.163 $\pm$ 0.025 & -0.275 $\pm$ 0.044 & 0.021 $\pm$ 0.017 & Cg\\
 & 0.913 $\pm$ 0.063 & 0.349 $\pm$ 0.028 & -0.633 $\pm$ 0.046 & 0.133 $\pm$ 0.003 & C8\\
 & 0.126 $\pm$ 0.035 & 0.481 $\pm$ 0.029 & -1.080 $\pm$ 0.031 & 0.061 $\pm$ 0.045 & B5\\
 & 0.429 $\pm$ 0.044 & 0.711 $\pm$ 0.040 & -1.694 $\pm$ 0.054 & 0.312 $\pm$ 0.008 & C7\\
 & 0.263 $\pm$ 0.034 & 0.213 $\pm$ 0.040 & -2.049 $\pm$ 0.054 & 0.312 $\pm$ 0.013 & C6\\
 & 0.116 $\pm$ 0.023 & 0.442 $\pm$ 0.029 & -2.385 $\pm$ 0.046 & 0.150 $\pm$ 0.027 & C5\\
 & 0.065 $\pm$ 0.017 & 1.257 $\pm$ 0.073 & -3.410 $\pm$ 0.082 & 0.690 $\pm$ 0.062 & C4\\
 & 0.013 $\pm$ 0.008 & 1.438 $\pm$ 0.054 & -4.553 $\pm$ 0.065 & 0.485 $\pm$ 0.245 & C3\\
 & 0.037 $\pm$ 0.014 & 2.601 $\pm$ 0.144 & -8.850 $\pm$ 0.148 & 1.419 $\pm$ 0.219 & C2\\
 & 0.021 $\pm$ 0.010 & 2.917 $\pm$ 0.129 & -11.161 $\pm$ 0.134 & 1.263 $\pm$ 0.312 & C1\\
\hline
1996.74 & 0.548 $\pm$ 0.072 & 0.000 $\pm$ 0.029 & 0.000 $\pm$ 0.030 & 0.048 $\pm$ 0.010 & CS\\
 & 1.056 $\pm$ 0.070 & 0.135 $\pm$ 0.030 & -0.291 $\pm$ 0.031 & 0.095 $\pm$ 0.003 & Cg\\
 & 0.838 $\pm$ 0.089 & 0.427 $\pm$ 0.031 & -0.776 $\pm$ 0.032 & 0.113 $\pm$ 0.007 & C8\\
 & 0.354 $\pm$ 0.046 & 0.791 $\pm$ 0.039 & -1.881 $\pm$ 0.040 & 0.266 $\pm$ 0.011 & C7\\
 & 0.338 $\pm$ 0.045 & 0.360 $\pm$ 0.044 & -2.334 $\pm$ 0.045 & 0.338 $\pm$ 0.012 & C6\\
 & 0.106 $\pm$ 0.033 & 1.146 $\pm$ 0.137 & -3.585 $\pm$ 0.138 & 1.343 $\pm$ 0.147 & C4\\
 & 0.065 $\pm$ 0.023 & 2.754 $\pm$ 0.169 & -9.167 $\pm$ 0.169 & 1.664 $\pm$ 0.211 & C2\\
 & 0.027 $\pm$ 0.015 & 2.842 $\pm$ 0.154 & -12.307 $\pm$ 0.154 & 1.510 $\pm$ 0.414 & C1\\
\hline
1996.82 & 0.482 $\pm$ 0.076 & 0.000 $\pm$ 0.022 & 0.000 $\pm$ 0.030 & 0.028 $\pm$ 0.003 & CS\\
 & 0.921 $\pm$ 0.062 & 0.077 $\pm$ 0.023 & -0.246 $\pm$ 0.031 & 0.070 $\pm$ 0.001 & Cg\\
 & 0.532 $\pm$ 0.024 & 0.353 $\pm$ 0.028 & -0.650 $\pm$ 0.034 & 0.170 $\pm$ 0.009 & C8\\
 & 0.539 $\pm$ 0.025 & 0.422 $\pm$ 0.024 & -0.864 $\pm$ 0.032 & 0.101 $\pm$ 0.008 & B6\\
 & 0.323 $\pm$ 0.036 & 0.792 $\pm$ 0.031 & -1.929 $\pm$ 0.037 & 0.221 $\pm$ 0.013 & C7\\
 & 0.149 $\pm$ 0.017 & 0.235 $\pm$ 0.034 & -2.224 $\pm$ 0.040 & 0.258 $\pm$ 0.002 & C6\\
 & 0.231 $\pm$ 0.022 & 0.456 $\pm$ 0.043 & -2.479 $\pm$ 0.047 & 0.368 $\pm$ 0.001 & C5\\
 & 0.041 $\pm$ 0.008 & 1.391 $\pm$ 0.050 & -3.581 $\pm$ 0.054 & 0.445 $\pm$ 0.003 & C4\\
 & 0.025 $\pm$ 0.007 & 1.470 $\pm$ 0.088 & -4.488 $\pm$ 0.090 & 0.849 $\pm$ 0.040 & C3\\
 & 0.025 $\pm$ 0.007 & 2.730 $\pm$ 0.088 & -8.647 $\pm$ 0.090 & 0.849 $\pm$ 0.041 & C2\\
 & 0.032 $\pm$ 0.018 & 3.183 $\pm$ 0.202 & -10.713 $\pm$ 0.203 & 2.004 $\pm$ 0.493 & C1\\
\hline
1996.93 & 1.080 $\pm$ 0.073 & 0.000 $\pm$ 0.026 & 0.000 $\pm$ 0.040 & 0.038 $\pm$ 0.003 & CS\\
 & 0.346 $\pm$ 0.034 & 0.193 $\pm$ 0.027 & -0.354 $\pm$ 0.040 & 0.061 $\pm$ 0.007 & Cg\\
 & 0.882 $\pm$ 0.035 & 0.372 $\pm$ 0.027 & -0.698 $\pm$ 0.041 & 0.095 $\pm$ 0.001 & C8\\
 & 0.339 $\pm$ 0.029 & 0.738 $\pm$ 0.037 & -1.743 $\pm$ 0.048 & 0.263 $\pm$ 0.006 & C7\\
 & 0.373 $\pm$ 0.047 & 0.320 $\pm$ 0.051 & -2.232 $\pm$ 0.060 & 0.444 $\pm$ 0.014 & C5\\
 & 0.039 $\pm$ 0.011 & 1.371 $\pm$ 0.040 & -3.645 $\pm$ 0.050 & 0.306 $\pm$ 0.063 & C4\\
 & 0.033 $\pm$ 0.007 & 1.734 $\pm$ 0.169 & -5.693 $\pm$ 0.172 & 1.674 $\pm$ 0.785 & C3\\
 & 0.055 $\pm$ 0.024 & 2.764 $\pm$ 0.151 & -11.212 $\pm$ 0.154 & 1.484 $\pm$ 0.294 & C1\\
\hline
1997.66 & 0.445 $\pm$ 0.059 & 0.000 $\pm$ 0.027 & 0.000 $\pm$ 0.038 & 0.035 $\pm$ 0.001 & CS\\
 & 0.915 $\pm$ 0.063 & 0.058 $\pm$ 0.027 & -0.219 $\pm$ 0.038 & 0.039 $\pm$ 0.001 & Cg\\
 & 0.229 $\pm$ 0.006 & 0.322 $\pm$ 0.032 & -0.629 $\pm$ 0.042 & 0.172 $\pm$ 0.012 & C9\\
 & 0.761 $\pm$ 0.014 & 0.466 $\pm$ 0.029 & -0.896 $\pm$ 0.039 & 0.111 $\pm$ 0.009 & C8\\
 & 0.220 $\pm$ 0.018 & 0.829 $\pm$ 0.045 & -2.114 $\pm$ 0.052 & 0.355 $\pm$ 0.001 & C7\\
 & 0.104 $\pm$ 0.014 & 0.266 $\pm$ 0.046 & -2.386 $\pm$ 0.053 & 0.367 $\pm$ 0.002 & C6\\
 & 0.213 $\pm$ 0.017 & 0.489 $\pm$ 0.052 & -2.734 $\pm$ 0.058 & 0.442 $\pm$ 0.001 & C5\\
 & 0.048 $\pm$ 0.010 & 1.353 $\pm$ 0.070 & -3.746 $\pm$ 0.075 & 0.643 $\pm$ 0.014 & C4\\
 & 0.032 $\pm$ 0.008 & 1.468 $\pm$ 0.117 & -5.075 $\pm$ 0.120 & 1.139 $\pm$ 0.062 & C3\\
 & 0.035 $\pm$ 0.007 & 2.653 $\pm$ 0.122 & -9.278 $\pm$ 0.125 & 1.194 $\pm$ 0.043 & C2\\
 & 0.052 $\pm$ 0.015 & 2.983 $\pm$ 0.168 & -11.270 $\pm$ 0.170 & 1.659 $\pm$ 0.126 & C1\\
\hline
1998.21 & 0.635 $\pm$ 0.045 & 0.000 $\pm$ 0.028 & 0.000 $\pm$ 0.041 & 0.072 $\pm$ 0.003 & CS\\
 & 1.071 $\pm$ 0.058 & 0.054 $\pm$ 0.031 & -0.239 $\pm$ 0.043 & 0.149 $\pm$ 0.002 & Cg\\
 & 0.367 $\pm$ 0.036 & 0.444 $\pm$ 0.036 & -0.812 $\pm$ 0.047 & 0.228 $\pm$ 0.006 & C9\\
 & 0.645 $\pm$ 0.047 & 0.492 $\pm$ 0.035 & -1.088 $\pm$ 0.046 & 0.212 $\pm$ 0.003 & C8\\
 & 0.144 $\pm$ 0.019 & 0.860 $\pm$ 0.054 & -2.214 $\pm$ 0.062 & 0.465 $\pm$ 0.012 & C7\\
 & 0.236 $\pm$ 0.024 & 0.465 $\pm$ 0.066 & -2.618 $\pm$ 0.072 & 0.599 $\pm$ 0.009 & C6\\
 & 0.058 $\pm$ 0.012 & 0.528 $\pm$ 0.046 & -3.198 $\pm$ 0.055 & 0.366 $\pm$ 0.027 & C5\\
 & 0.049 $\pm$ 0.011 & 1.384 $\pm$ 0.075 & -3.908 $\pm$ 0.081 & 0.703 $\pm$ 0.048 & C4\\
 & 0.021 $\pm$ 0.008 & 1.620 $\pm$ 0.081 & -5.565 $\pm$ 0.087 & 0.767 $\pm$ 0.123 & C3\\
 & 0.047 $\pm$ 0.019 & 2.687 $\pm$ 0.181 & -9.720 $\pm$ 0.184 & 1.793 $\pm$ 0.273 & C2\\
 & 0.044 $\pm$ 0.022 & 2.743 $\pm$ 0.225 & -12.199 $\pm$ 0.227 & 2.232 $\pm$ 0.492 & C1\\
\hline
1998.59 & 0.737 $\pm$ 0.046 & 0.000 $\pm$ 0.029 & 0.000 $\pm$ 0.036 & 0.088 $\pm$ 0.002 & CS\\
 & 0.923 $\pm$ 0.052 & 0.085 $\pm$ 0.028 & -0.261 $\pm$ 0.036 & 0.057 $\pm$ 0.001 & Cg\\
 & 0.216 $\pm$ 0.025 & 0.262 $\pm$ 0.034 & -0.586 $\pm$ 0.041 & 0.205 $\pm$ 0.006 & C10\\
 & 0.424 $\pm$ 0.035 & 0.510 $\pm$ 0.030 & -0.937 $\pm$ 0.038 & 0.134 $\pm$ 0.003 & C9\\
 & 0.481 $\pm$ 0.037 & 0.547 $\pm$ 0.039 & -1.204 $\pm$ 0.045 & 0.275 $\pm$ 0.003 & C8\\
 & 0.118 $\pm$ 0.016 & 0.938 $\pm$ 0.049 & -2.276 $\pm$ 0.054 & 0.410 $\pm$ 0.010 & C7\\
 & 0.245 $\pm$ 0.024 & 0.496 $\pm$ 0.066 & -2.660 $\pm$ 0.070 & 0.601 $\pm$ 0.006 & C6\\
 & 0.084 $\pm$ 0.014 & 0.573 $\pm$ 0.049 & -3.287 $\pm$ 0.054 & 0.410 $\pm$ 0.014 & C5\\
 & 0.044 $\pm$ 0.010 & 1.385 $\pm$ 0.066 & -3.780 $\pm$ 0.070 & 0.606 $\pm$ 0.036 & C4\\
 & 0.039 $\pm$ 0.012 & 1.554 $\pm$ 0.123 & -5.336 $\pm$ 0.125 & 1.204 $\pm$ 0.123 & C3\\
 & 0.042 $\pm$ 0.013 & 2.642 $\pm$ 0.151 & -9.580 $\pm$ 0.152 & 1.483 $\pm$ 0.150 & C2\\
 & 0.049 $\pm$ 0.017 & 2.893 $\pm$ 0.193 & -11.840 $\pm$ 0.194 & 1.908 $\pm$ 0.229 & C1\\
\hline
1999.01 & 0.588 $\pm$ 0.041 & 0.000 $\pm$ 0.030 & 0.000 $\pm$ 0.034 & 0.110 $\pm$ 0.002 & CS\\
 & 0.675 $\pm$ 0.044 & 0.050 $\pm$ 0.029 & -0.185 $\pm$ 0.032 & 0.060 $\pm$ 0.001 & Cg\\
 & 0.322 $\pm$ 0.030 & 0.166 $\pm$ 0.035 & -0.469 $\pm$ 0.038 & 0.209 $\pm$ 0.003 & C10\\
 & 0.228 $\pm$ 0.026 & 0.501 $\pm$ 0.035 & -0.887 $\pm$ 0.037 & 0.200 $\pm$ 0.005 & C9\\
 & 0.435 $\pm$ 0.035 & 0.575 $\pm$ 0.042 & -1.239 $\pm$ 0.044 & 0.308 $\pm$ 0.003 & C8\\
 & 0.098 $\pm$ 0.017 & 0.966 $\pm$ 0.058 & -2.339 $\pm$ 0.060 & 0.509 $\pm$ 0.016 & C7\\
 & 0.185 $\pm$ 0.023 & 0.478 $\pm$ 0.073 & -2.749 $\pm$ 0.075 & 0.675 $\pm$ 0.011 & C6\\
 & 0.054 $\pm$ 0.013 & 0.575 $\pm$ 0.040 & -3.394 $\pm$ 0.043 & 0.284 $\pm$ 0.021 & C5\\
 & 0.065 $\pm$ 0.015 & 1.228 $\pm$ 0.104 & -3.775 $\pm$ 0.105 & 0.998 $\pm$ 0.055 & C4\\
 & 0.022 $\pm$ 0.007 & 1.644 $\pm$ 0.074 & -5.766 $\pm$ 0.075 & 0.681 $\pm$ 0.068 & C3\\
 & 0.034 $\pm$ 0.012 & 2.800 $\pm$ 0.135 & -9.789 $\pm$ 0.135 & 1.315 $\pm$ 0.149 & C2\\
 & 0.038 $\pm$ 0.015 & 2.714 $\pm$ 0.176 & -11.921 $\pm$ 0.176 & 1.733 $\pm$ 0.249 & C1\\
\hline
1999.54 & 0.606 $\pm$ 0.036 & 0.000 $\pm$ 0.033 & 0.000 $\pm$ 0.033 & 0.065 $\pm$ 0.002 & CS\\
 & 1.061 $\pm$ 0.048 & 0.066 $\pm$ 0.033 & -0.216 $\pm$ 0.032 & 0.040 $\pm$ 0.001 & Cg\\
 & 0.467 $\pm$ 0.032 & 0.226 $\pm$ 0.036 & -0.578 $\pm$ 0.036 & 0.155 $\pm$ 0.002 & C10\\
 & 0.162 $\pm$ 0.019 & 0.564 $\pm$ 0.034 & -0.981 $\pm$ 0.034 & 0.109 $\pm$ 0.006 & C9\\
 & 0.435 $\pm$ 0.031 & 0.623 $\pm$ 0.046 & -1.363 $\pm$ 0.046 & 0.327 $\pm$ 0.003 & C8\\
 & 0.118 $\pm$ 0.016 & 1.029 $\pm$ 0.053 & -2.529 $\pm$ 0.052 & 0.411 $\pm$ 0.011 & C7\\
 & 0.130 $\pm$ 0.017 & 0.461 $\pm$ 0.065 & -2.847 $\pm$ 0.065 & 0.563 $\pm$ 0.012 & C6\\
 & 0.095 $\pm$ 0.015 & 0.551 $\pm$ 0.051 & -3.524 $\pm$ 0.050 & 0.386 $\pm$ 0.013 & C5\\
 & 0.073 $\pm$ 0.014 & 1.243 $\pm$ 0.095 & -3.937 $\pm$ 0.095 & 0.895 $\pm$ 0.036 & C4\\
 & 0.030 $\pm$ 0.007 & 1.647 $\pm$ 0.120 & -5.831 $\pm$ 0.119 & 1.150 $\pm$ 0.069 & C3\\
 & 0.041 $\pm$ 0.010 & 2.722 $\pm$ 0.157 & -9.961 $\pm$ 0.157 & 1.537 $\pm$ 0.090 & C2\\
 & 0.046 $\pm$ 0.011 & 2.689 $\pm$ 0.167 & -11.746 $\pm$ 0.167 & 1.637 $\pm$ 0.092 & C1\\
\hline
1999.56 & 0.525 $\pm$ 0.044 & 0.000 $\pm$ 0.034 & 0.000 $\pm$ 0.033 & 0.050 $\pm$ 0.003 & CS\\
 & 1.123 $\pm$ 0.064 & 0.064 $\pm$ 0.034 & -0.220 $\pm$ 0.033 & 0.061 $\pm$ 0.002 & Cg\\
 & 0.474 $\pm$ 0.042 & 0.221 $\pm$ 0.037 & -0.585 $\pm$ 0.036 & 0.157 $\pm$ 0.004 & C10\\
 & 0.177 $\pm$ 0.026 & 0.554 $\pm$ 0.038 & -0.996 $\pm$ 0.038 & 0.186 $\pm$ 0.010 & C9\\
 & 0.431 $\pm$ 0.040 & 0.626 $\pm$ 0.046 & -1.385 $\pm$ 0.046 & 0.318 $\pm$ 0.005 & C8\\
 & 0.106 $\pm$ 0.019 & 1.045 $\pm$ 0.050 & -2.550 $\pm$ 0.049 & 0.372 $\pm$ 0.018 & C7\\
 & 0.147 $\pm$ 0.023 & 0.495 $\pm$ 0.069 & -2.874 $\pm$ 0.069 & 0.607 $\pm$ 0.017 & C6\\
 & 0.092 $\pm$ 0.018 & 0.544 $\pm$ 0.053 & -3.575 $\pm$ 0.052 & 0.409 $\pm$ 0.021 & C5\\
 & 0.069 $\pm$ 0.017 & 1.294 $\pm$ 0.098 & -4.013 $\pm$ 0.098 & 0.920 $\pm$ 0.060 & C4\\
 & 0.023 $\pm$ 0.006 & 1.728 $\pm$ 0.082 & -6.048 $\pm$ 0.081 & 0.743 $\pm$ 0.061 & C3\\
 & 0.049 $\pm$ 0.013 & 2.876 $\pm$ 0.167 & -10.111 $\pm$ 0.167 & 1.640 $\pm$ 0.113 & C2\\
 & 0.043 $\pm$ 0.013 & 2.705 $\pm$ 0.189 & -11.944 $\pm$ 0.189 & 1.861 $\pm$ 0.173 & C1\\
\hline
1999.57 & 0.614 $\pm$ 0.063 & 0.000 $\pm$ 0.029 & 0.000 $\pm$ 0.058 & 0.040 $\pm$ 0.012 & CS\\
 & 1.006 $\pm$ 0.102 & 0.062 $\pm$ 0.029 & -0.209 $\pm$ 0.058 & 0.048 $\pm$ 0.012 & Cg\\
 & 0.525 $\pm$ 0.075 & 0.219 $\pm$ 0.034 & -0.544 $\pm$ 0.061 & 0.187 $\pm$ 0.024 & C10\\
 & 0.265 $\pm$ 0.045 & 0.559 $\pm$ 0.041 & -1.075 $\pm$ 0.065 & 0.299 $\pm$ 0.035 & C9\\
 & 0.340 $\pm$ 0.051 & 0.630 $\pm$ 0.040 & -1.423 $\pm$ 0.064 & 0.280 $\pm$ 0.027 & C8\\
 & 0.114 $\pm$ 0.020 & 1.011 $\pm$ 0.038 & -2.607 $\pm$ 0.063 & 0.254 $\pm$ 0.035 & C7\\
 & 0.109 $\pm$ 0.026 & 0.434 $\pm$ 0.045 & -2.852 $\pm$ 0.067 & 0.344 $\pm$ 0.066 & C6\\
 & 0.112 $\pm$ 0.026 & 0.581 $\pm$ 0.050 & -3.531 $\pm$ 0.071 & 0.416 $\pm$ 0.067 & C5\\
 & 0.050 $\pm$ 0.019 & 1.231 $\pm$ 0.070 & -4.067 $\pm$ 0.086 & 0.641 $\pm$ 0.175 & C4\\
 & 0.037 $\pm$ 0.016 & 1.596 $\pm$ 0.089 & -5.517 $\pm$ 0.103 & 0.849 $\pm$ 0.244 & C3\\
 & 0.010 $\pm$ 0.007 & 2.959 $\pm$ 0.055 & -9.485 $\pm$ 0.075 & 0.476 $\pm$ 0.449 & C2\\
 & 0.065 $\pm$ 0.024 & 2.767 $\pm$ 0.192 & -11.447 $\pm$ 0.199 & 1.903 $\pm$ 0.280 & C1\\
\hline
1999.79 & 0.441 $\pm$ 0.039 & 0.000 $\pm$ 0.030 & 0.000 $\pm$ 0.030 & 0.056 $\pm$ 0.003 & CS\\
 & 0.848 $\pm$ 0.055 & 0.063 $\pm$ 0.031 & -0.228 $\pm$ 0.031 & 0.086 $\pm$ 0.001 & Cg\\
 & 0.605 $\pm$ 0.046 & 0.225 $\pm$ 0.036 & -0.593 $\pm$ 0.036 & 0.209 $\pm$ 0.002 & C10\\
 & 0.171 $\pm$ 0.024 & 0.584 $\pm$ 0.045 & -1.063 $\pm$ 0.044 & 0.331 $\pm$ 0.009 & C9\\
 & 0.353 $\pm$ 0.035 & 0.636 $\pm$ 0.045 & -1.459 $\pm$ 0.044 & 0.332 $\pm$ 0.004 & C8\\
 & 0.100 $\pm$ 0.017 & 1.067 $\pm$ 0.039 & -2.636 $\pm$ 0.038 & 0.248 $\pm$ 0.011 & C7\\
 & 0.124 $\pm$ 0.019 & 0.501 $\pm$ 0.065 & -2.913 $\pm$ 0.064 & 0.572 $\pm$ 0.013 & C6\\
 & 0.105 $\pm$ 0.017 & 0.571 $\pm$ 0.062 & -3.623 $\pm$ 0.061 & 0.539 $\pm$ 0.015 & C5\\
 & 0.064 $\pm$ 0.014 & 1.355 $\pm$ 0.091 & -4.131 $\pm$ 0.091 & 0.857 $\pm$ 0.042 & C4\\
 & 0.022 $\pm$ 0.008 & 1.435 $\pm$ 0.056 & -6.607 $\pm$ 0.056 & 0.474 $\pm$ 0.064 & C3\\
 & 0.025 $\pm$ 0.009 & 2.546 $\pm$ 0.085 & -9.594 $\pm$ 0.085 & 0.793 $\pm$ 0.096 & C2\\
 & 0.051 $\pm$ 0.013 & 2.728 $\pm$ 0.142 & -11.215 $\pm$ 0.142 & 1.390 $\pm$ 0.097 & C1\\
\hline
1999.98 & 0.373 $\pm$ 0.056 & 0.000 $\pm$ 0.031 & 0.000 $\pm$ 0.029 & 0.052 $\pm$ 0.008 & CS\\
 & 0.997 $\pm$ 0.091 & 0.059 $\pm$ 0.031 & -0.251 $\pm$ 0.029 & 0.057 $\pm$ 0.003 & Cg\\
 & 0.698 $\pm$ 0.076 & 0.235 $\pm$ 0.036 & -0.649 $\pm$ 0.035 & 0.196 $\pm$ 0.004 & C10\\
 & 0.175 $\pm$ 0.028 & 0.615 $\pm$ 0.044 & -1.193 $\pm$ 0.043 & 0.318 $\pm$ 0.010 & C9\\
 & 0.287 $\pm$ 0.036 & 0.656 $\pm$ 0.042 & -1.572 $\pm$ 0.041 & 0.292 $\pm$ 0.006 & C8\\
 & 0.134 $\pm$ 0.024 & 1.082 $\pm$ 0.043 & -2.699 $\pm$ 0.041 & 0.297 $\pm$ 0.013 & C7\\
 & 0.143 $\pm$ 0.026 & 0.477 $\pm$ 0.080 & -3.088 $\pm$ 0.079 & 0.740 $\pm$ 0.025 & C6\\
 & 0.058 $\pm$ 0.016 & 0.509 $\pm$ 0.051 & -3.764 $\pm$ 0.050 & 0.406 $\pm$ 0.035 & C5\\
 & 0.075 $\pm$ 0.020 & 1.158 $\pm$ 0.097 & -3.998 $\pm$ 0.097 & 0.922 $\pm$ 0.065 & C4\\
 & 0.014 $\pm$ 0.009 & 1.539 $\pm$ 0.063 & -6.354 $\pm$ 0.062 & 0.554 $\pm$ 0.184 & C3\\
 & 0.022 $\pm$ 0.010 & 3.111 $\pm$ 0.132 & -9.531 $\pm$ 0.132 & 1.284 $\pm$ 0.256 & C2\\
 & 0.048 $\pm$ 0.015 & 2.783 $\pm$ 0.141 & -11.243 $\pm$ 0.141 & 1.377 $\pm$ 0.136 & C1\\
\hline
2000.01 & 0.493 $\pm$ 0.045 & 0.000 $\pm$ 0.029 & 0.000 $\pm$ 0.035 & 0.043 $\pm$ 0.004 & CS\\
 & 0.814 $\pm$ 0.057 & 0.064 $\pm$ 0.029 & -0.233 $\pm$ 0.035 & 0.056 $\pm$ 0.002 & Cg\\
 & 0.624 $\pm$ 0.054 & 0.238 $\pm$ 0.032 & -0.622 $\pm$ 0.038 & 0.151 $\pm$ 0.003 & C10\\
 & 0.135 $\pm$ 0.025 & 0.613 $\pm$ 0.035 & -1.118 $\pm$ 0.040 & 0.205 $\pm$ 0.016 & C9\\
 & 0.312 $\pm$ 0.038 & 0.658 $\pm$ 0.044 & -1.510 $\pm$ 0.049 & 0.343 $\pm$ 0.008 & C8\\
 & 0.126 $\pm$ 0.017 & 1.084 $\pm$ 0.038 & -2.654 $\pm$ 0.043 & 0.258 $\pm$ 0.008 & C7\\
 & 0.115 $\pm$ 0.016 & 0.507 $\pm$ 0.070 & -3.035 $\pm$ 0.073 & 0.642 $\pm$ 0.015 & C6\\
 & 0.076 $\pm$ 0.013 & 0.545 $\pm$ 0.047 & -3.691 $\pm$ 0.052 & 0.380 $\pm$ 0.016 & C5\\
 & 0.065 $\pm$ 0.013 & 1.206 $\pm$ 0.093 & -4.131 $\pm$ 0.095 & 0.883 $\pm$ 0.039 & C4\\
 & 0.021 $\pm$ 0.006 & 1.680 $\pm$ 0.109 & -5.977 $\pm$ 0.111 & 1.051 $\pm$ 0.095 & C3\\
 & 0.032 $\pm$ 0.009 & 2.858 $\pm$ 0.160 & -10.245 $\pm$ 0.161 & 1.572 $\pm$ 0.117 & C2\\
 & 0.038 $\pm$ 0.009 & 2.627 $\pm$ 0.156 & -11.892 $\pm$ 0.158 & 1.537 $\pm$ 0.094 & C1\\
\hline
2000.35 & 0.474 $\pm$ 0.043 & 0.000 $\pm$ 0.029 & 0.000 $\pm$ 0.035 & 0.019 $\pm$ 0.003 & CS\\
 & 0.790 $\pm$ 0.055 & 0.103 $\pm$ 0.030 & -0.281 $\pm$ 0.036 & 0.074 $\pm$ 0.002 & Cg\\
 & 0.751 $\pm$ 0.076 & 0.265 $\pm$ 0.032 & -0.672 $\pm$ 0.037 & 0.135 $\pm$ 0.005 & C10\\
 & 0.145 $\pm$ 0.034 & 0.464 $\pm$ 0.034 & -0.941 $\pm$ 0.039 & 0.169 $\pm$ 0.024 & C9\\
 & 0.385 $\pm$ 0.027 & 0.706 $\pm$ 0.049 & -1.573 $\pm$ 0.052 & 0.388 $\pm$ 0.003 & C8\\
 & 0.176 $\pm$ 0.037 & 1.114 $\pm$ 0.041 & -2.796 $\pm$ 0.045 & 0.290 $\pm$ 0.022 & C7\\
 & 0.147 $\pm$ 0.035 & 0.541 $\pm$ 0.069 & -3.268 $\pm$ 0.072 & 0.630 $\pm$ 0.040 & C6\\
 & 0.062 $\pm$ 0.023 & 0.609 $\pm$ 0.043 & -3.943 $\pm$ 0.047 & 0.312 $\pm$ 0.063 & C5\\
 & 0.073 $\pm$ 0.027 & 1.240 $\pm$ 0.092 & -4.310 $\pm$ 0.094 & 0.870 $\pm$ 0.119 & C4\\
 & 0.022 $\pm$ 0.016 & 1.698 $\pm$ 0.095 & -6.192 $\pm$ 0.097 & 0.905 $\pm$ 0.419 & C3\\
 & 0.029 $\pm$ 0.008 & 2.928 $\pm$ 0.142 & -10.247 $\pm$ 0.144 & 1.393 $\pm$ 0.097 & C2\\
 & 0.052 $\pm$ 0.011 & 2.702 $\pm$ 0.168 & -11.987 $\pm$ 0.170 & 1.659 $\pm$ 0.079 & C1\\
\hline
2001.17 & 0.344 $\pm$ 0.037 & 0.000 $\pm$ 0.028 & 0.000 $\pm$ 0.040 & 0.070 $\pm$ 0.006 & CS\\
 & 0.740 $\pm$ 0.054 & 0.055 $\pm$ 0.028 & -0.179 $\pm$ 0.040 & 0.067 $\pm$ 0.003 & Cg\\
 & 0.367 $\pm$ 0.038 & 0.254 $\pm$ 0.030 & -0.657 $\pm$ 0.042 & 0.133 $\pm$ 0.005 & C11\\
 & 0.592 $\pm$ 0.048 & 0.409 $\pm$ 0.032 & -0.929 $\pm$ 0.043 & 0.177 $\pm$ 0.003 & C10\\
 & 0.250 $\pm$ 0.030 & 0.796 $\pm$ 0.052 & -1.820 $\pm$ 0.060 & 0.447 $\pm$ 0.009 & C8\\
 & 0.127 $\pm$ 0.022 & 1.134 $\pm$ 0.046 & -3.053 $\pm$ 0.055 & 0.377 $\pm$ 0.017 & C7\\
 & 0.103 $\pm$ 0.018 & 0.581 $\pm$ 0.065 & -3.683 $\pm$ 0.072 & 0.597 $\pm$ 0.022 & C6\\
 & 0.035 $\pm$ 0.009 & 0.789 $\pm$ 0.038 & -4.387 $\pm$ 0.048 & 0.268 $\pm$ 0.034 & C5\\
 & 0.054 $\pm$ 0.012 & 1.314 $\pm$ 0.099 & -4.633 $\pm$ 0.103 & 0.954 $\pm$ 0.055 & C4\\
 & 0.015 $\pm$ 0.007 & 1.614 $\pm$ 0.084 & -6.500 $\pm$ 0.089 & 0.796 $\pm$ 0.180 & C3\\
 & 0.019 $\pm$ 0.007 & 2.780 $\pm$ 0.144 & -11.019 $\pm$ 0.147 & 1.415 $\pm$ 0.367 & C2\\
 & 0.040 $\pm$ 0.013 & 2.533 $\pm$ 0.128 & -12.324 $\pm$ 0.131 & 1.248 $\pm$ 0.136 & C1\\
\hline
2001.38 & 0.634 $\pm$ 0.044 & 0.000 $\pm$ 0.027 & 0.000 $\pm$ 0.036 & 0.021 $\pm$ 0.002 & CS\\
 & 0.721 $\pm$ 0.047 & 0.066 $\pm$ 0.028 & -0.192 $\pm$ 0.037 & 0.073 $\pm$ 0.002 & Cg\\
 & 0.498 $\pm$ 0.039 & 0.275 $\pm$ 0.032 & -0.688 $\pm$ 0.040 & 0.163 $\pm$ 0.003 & C11\\
 & 0.554 $\pm$ 0.041 & 0.426 $\pm$ 0.033 & -0.953 $\pm$ 0.041 & 0.181 $\pm$ 0.003 & C10\\
 & 0.281 $\pm$ 0.031 & 0.800 $\pm$ 0.057 & -1.846 $\pm$ 0.062 & 0.506 $\pm$ 0.008 & C8\\
 & 0.136 $\pm$ 0.029 & 1.123 $\pm$ 0.051 & -3.088 $\pm$ 0.057 & 0.434 $\pm$ 0.025 & C7\\
 & 0.152 $\pm$ 0.024 & 0.618 $\pm$ 0.081 & -3.872 $\pm$ 0.085 & 0.765 $\pm$ 0.021 & C6\\
 & 0.069 $\pm$ 0.015 & 1.261 $\pm$ 0.082 & -4.760 $\pm$ 0.086 & 0.777 $\pm$ 0.039 & C4\\
 & 0.019 $\pm$ 0.009 & 1.943 $\pm$ 0.101 & -6.630 $\pm$ 0.104 & 0.970 $\pm$ 0.196 & C3\\
 & 0.083 $\pm$ 0.031 & 2.726 $\pm$ 0.215 & -11.953 $\pm$ 0.216 & 2.128 $\pm$ 0.288 & C1\\
\hline
2001.50 & 0.703 $\pm$ 0.047 & 0.000 $\pm$ 0.028 & 0.000 $\pm$ 0.037 & 0.040 $\pm$ 0.002 & CS\\
 & 0.708 $\pm$ 0.047 & 0.061 $\pm$ 0.028 & -0.177 $\pm$ 0.037 & 0.060 $\pm$ 0.002 & Cg\\
 & 0.470 $\pm$ 0.053 & 0.279 $\pm$ 0.033 & -0.686 $\pm$ 0.041 & 0.180 $\pm$ 0.006 & C11\\
 & 0.545 $\pm$ 0.057 & 0.425 $\pm$ 0.033 & -0.949 $\pm$ 0.040 & 0.174 $\pm$ 0.005 & C10\\
 & 0.276 $\pm$ 0.032 & 0.800 $\pm$ 0.058 & -1.850 $\pm$ 0.063 & 0.515 $\pm$ 0.009 & C8\\
 & 0.119 $\pm$ 0.024 & 1.126 $\pm$ 0.047 & -3.072 $\pm$ 0.053 & 0.381 $\pm$ 0.022 & C7\\
 & 0.127 $\pm$ 0.023 & 0.599 $\pm$ 0.079 & -3.742 $\pm$ 0.082 & 0.738 $\pm$ 0.027 & C6\\
 & 0.105 $\pm$ 0.025 & 1.122 $\pm$ 0.102 & -4.575 $\pm$ 0.105 & 0.980 $\pm$ 0.060 & C4\\
 & 0.015 $\pm$ 0.007 & 1.772 $\pm$ 0.061 & -6.451 $\pm$ 0.066 & 0.549 $\pm$ 0.117 & C3\\
 & 0.016 $\pm$ 0.007 & 3.042 $\pm$ 0.113 & -10.374 $\pm$ 0.116 & 1.100 $\pm$ 0.198 & C2\\
 & 0.059 $\pm$ 0.015 & 2.687 $\pm$ 0.154 & -12.243 $\pm$ 0.156 & 1.517 $\pm$ 0.101 & C1\\
\hline
2001.84 & 0.618 $\pm$ 0.052 & 0.000 $\pm$ 0.028 & 0.000 $\pm$ 0.036 & 0.065 $\pm$ 0.003 & CS\\
 & 1.102 $\pm$ 0.069 & 0.051 $\pm$ 0.027 & -0.163 $\pm$ 0.036 & 0.051 $\pm$ 0.002 & Cg\\
 & 0.423 $\pm$ 0.052 & 0.282 $\pm$ 0.030 & -0.681 $\pm$ 0.038 & 0.140 $\pm$ 0.007 & C11\\
 & 0.560 $\pm$ 0.060 & 0.432 $\pm$ 0.033 & -0.996 $\pm$ 0.041 & 0.197 $\pm$ 0.005 & C10\\
 & 0.242 $\pm$ 0.031 & 0.830 $\pm$ 0.061 & -1.939 $\pm$ 0.066 & 0.550 $\pm$ 0.011 & C8\\
 & 0.123 $\pm$ 0.022 & 1.080 $\pm$ 0.056 & -3.215 $\pm$ 0.061 & 0.488 $\pm$ 0.020 & C7\\
 & 0.113 $\pm$ 0.022 & 0.617 $\pm$ 0.082 & -3.979 $\pm$ 0.085 & 0.772 $\pm$ 0.033 & C6\\
 & 0.080 $\pm$ 0.020 & 1.269 $\pm$ 0.101 & -4.797 $\pm$ 0.104 & 0.974 $\pm$ 0.064 & C4\\
 & 0.009 $\pm$ 0.008 & 1.869 $\pm$ 0.096 & -6.464 $\pm$ 0.098 & 0.917 $\pm$ 0.544 & C3\\
 & 0.015 $\pm$ 0.007 & 2.778 $\pm$ 0.117 & -10.696 $\pm$ 0.120 & 1.143 $\pm$ 0.263 & C2\\
 & 0.057 $\pm$ 0.016 & 2.722 $\pm$ 0.149 & -12.302 $\pm$ 0.151 & 1.463 $\pm$ 0.113 & C1\\
\hline
2001.97 & 0.636 $\pm$ 0.040 & 0.000 $\pm$ 0.027 & 0.000 $\pm$ 0.039 & 0.020 $\pm$ 0.002 & CS\\
 & 1.033 $\pm$ 0.051 & 0.051 $\pm$ 0.027 & -0.166 $\pm$ 0.039 & 0.054 $\pm$ 0.001 & Cg\\
 & 0.380 $\pm$ 0.044 & 0.279 $\pm$ 0.030 & -0.697 $\pm$ 0.041 & 0.126 $\pm$ 0.007 & C11\\
 & 0.483 $\pm$ 0.050 & 0.443 $\pm$ 0.034 & -1.023 $\pm$ 0.044 & 0.210 $\pm$ 0.006 & C10\\
 & 0.207 $\pm$ 0.034 & 0.841 $\pm$ 0.061 & -1.999 $\pm$ 0.067 & 0.545 $\pm$ 0.019 & C8\\
 & 0.103 $\pm$ 0.018 & 1.078 $\pm$ 0.059 & -3.203 $\pm$ 0.066 & 0.529 $\pm$ 0.021 & C7\\
 & 0.124 $\pm$ 0.022 & 0.687 $\pm$ 0.087 & -4.083 $\pm$ 0.091 & 0.824 $\pm$ 0.030 & C6\\
 & 0.050 $\pm$ 0.013 & 1.447 $\pm$ 0.080 & -4.996 $\pm$ 0.085 & 0.754 $\pm$ 0.059 & C4\\
 & 0.009 $\pm$ 0.007 & 1.616 $\pm$ 0.090 & -6.729 $\pm$ 0.095 & 0.863 $\pm$ 0.427 & C3\\
 & 0.061 $\pm$ 0.020 & 2.832 $\pm$ 0.185 & -12.057 $\pm$ 0.187 & 1.827 $\pm$ 0.194 & C1\\
\hline
2002.02 & 0.588 $\pm$ 0.051 & 0.000 $\pm$ 0.030 & 0.000 $\pm$ 0.038 & 0.031 $\pm$ 0.004 & CS\\
 & 1.373 $\pm$ 0.078 & 0.050 $\pm$ 0.030 & -0.168 $\pm$ 0.038 & 0.013 $\pm$ 0.002 & Cg\\
 & 0.403 $\pm$ 0.046 & 0.291 $\pm$ 0.034 & -0.694 $\pm$ 0.041 & 0.169 $\pm$ 0.007 & C11\\
 & 0.531 $\pm$ 0.053 & 0.439 $\pm$ 0.037 & -1.039 $\pm$ 0.043 & 0.215 $\pm$ 0.006 & C10\\
 & 0.231 $\pm$ 0.034 & 0.835 $\pm$ 0.064 & -1.976 $\pm$ 0.068 & 0.570 $\pm$ 0.017 & C8\\
 & 0.141 $\pm$ 0.027 & 1.037 $\pm$ 0.061 & -3.280 $\pm$ 0.065 & 0.533 $\pm$ 0.027 & C7\\
 & 0.114 $\pm$ 0.024 & 0.690 $\pm$ 0.087 & -4.260 $\pm$ 0.090 & 0.815 $\pm$ 0.044 & C6\\
 & 0.062 $\pm$ 0.020 & 1.477 $\pm$ 0.096 & -4.958 $\pm$ 0.099 & 0.917 $\pm$ 0.105 & C4\\
 & 0.085 $\pm$ 0.025 & 2.682 $\pm$ 0.204 & -12.245 $\pm$ 0.206 & 2.022 $\pm$ 0.180 & C1\\
\hline
2002.13 & 0.560 $\pm$ 0.050 & 0.000 $\pm$ 0.027 & 0.000 $\pm$ 0.041 & 0.011 $\pm$ 0.004 & CS\\
 & 1.403 $\pm$ 0.078 & 0.046 $\pm$ 0.028 & -0.161 $\pm$ 0.041 & 0.035 $\pm$ 0.002 & Cg\\
 & 0.340 $\pm$ 0.039 & 0.283 $\pm$ 0.030 & -0.695 $\pm$ 0.043 & 0.112 $\pm$ 0.007 & C11\\
 & 0.476 $\pm$ 0.046 & 0.444 $\pm$ 0.034 & -1.044 $\pm$ 0.046 & 0.204 $\pm$ 0.005 & C10\\
 & 0.202 $\pm$ 0.031 & 0.849 $\pm$ 0.063 & -1.992 $\pm$ 0.070 & 0.568 $\pm$ 0.018 & C8\\
 & 0.111 $\pm$ 0.023 & 1.023 $\pm$ 0.060 & -3.274 $\pm$ 0.067 & 0.529 $\pm$ 0.032 & C7\\
 & 0.132 $\pm$ 0.028 & 0.763 $\pm$ 0.100 & -4.262 $\pm$ 0.105 & 0.965 $\pm$ 0.050 & C6\\
 & 0.032 $\pm$ 0.013 & 1.588 $\pm$ 0.058 & -5.306 $\pm$ 0.066 & 0.515 $\pm$ 0.109 & C4\\
 & 0.064 $\pm$ 0.028 & 2.891 $\pm$ 0.181 & -12.187 $\pm$ 0.184 & 1.791 $\pm$ 0.333 & C1\\
\hline
2002.45 & 0.473 $\pm$ 0.036 & 0.000 $\pm$ 0.031 & 0.000 $\pm$ 0.028 & 0.051 $\pm$ 0.002 & CS\\
 & 2.061 $\pm$ 0.074 & 0.066 $\pm$ 0.031 & -0.205 $\pm$ 0.028 & 0.065 $\pm$ 0.000 & Cg\\
 & 0.271 $\pm$ 0.034 & 0.299 $\pm$ 0.032 & -0.729 $\pm$ 0.029 & 0.099 $\pm$ 0.005 & C11\\
 & 0.445 $\pm$ 0.032 & 0.456 $\pm$ 0.038 & -1.100 $\pm$ 0.035 & 0.220 $\pm$ 0.002 & C10\\
 & 0.210 $\pm$ 0.030 & 0.874 $\pm$ 0.073 & -2.160 $\pm$ 0.072 & 0.665 $\pm$ 0.014 & C8\\
 & 0.124 $\pm$ 0.020 & 0.979 $\pm$ 0.065 & -3.449 $\pm$ 0.063 & 0.570 $\pm$ 0.015 & C7\\
 & 0.094 $\pm$ 0.016 & 0.720 $\pm$ 0.091 & -4.378 $\pm$ 0.090 & 0.861 $\pm$ 0.025 & C6\\
 & 0.056 $\pm$ 0.014 & 1.500 $\pm$ 0.085 & -5.071 $\pm$ 0.084 & 0.792 $\pm$ 0.050 & C4\\
 & 0.015 $\pm$ 0.008 & 1.786 $\pm$ 0.150 & -7.156 $\pm$ 0.150 & 1.473 $\pm$ 0.365 & C3\\
 & 0.067 $\pm$ 0.020 & 2.611 $\pm$ 0.183 & -12.321 $\pm$ 0.183 & 1.808 $\pm$ 0.156 & C1\\
\hline
2003.66 & 0.338 $\pm$ 0.033 & 0.000 $\pm$ 0.032 & 0.000 $\pm$ 0.033 & 0.018 $\pm$ 0.004 & CS\\
 & 1.356 $\pm$ 0.066 & 0.078 $\pm$ 0.033 & -0.274 $\pm$ 0.034 & 0.070 $\pm$ 0.001 & Cg\\
 & 0.960 $\pm$ 0.056 & 0.177 $\pm$ 0.035 & -0.523 $\pm$ 0.036 & 0.147 $\pm$ 0.002 & C12\\
 & 0.163 $\pm$ 0.032 & 0.374 $\pm$ 0.037 & -0.888 $\pm$ 0.038 & 0.187 $\pm$ 0.018 & C11\\
 & 0.226 $\pm$ 0.038 & 0.592 $\pm$ 0.050 & -1.431 $\pm$ 0.051 & 0.388 $\pm$ 0.016 & C10\\
 & 0.132 $\pm$ 0.030 & 0.993 $\pm$ 0.076 & -2.446 $\pm$ 0.077 & 0.693 $\pm$ 0.041 & C8\\
 & 0.079 $\pm$ 0.019 & 1.054 $\pm$ 0.079 & -3.478 $\pm$ 0.079 & 0.724 $\pm$ 0.044 & C7\\
 & 0.060 $\pm$ 0.014 & 0.848 $\pm$ 0.092 & -4.397 $\pm$ 0.093 & 0.867 $\pm$ 0.052 & C6\\
 & 0.060 $\pm$ 0.013 & 1.507 $\pm$ 0.113 & -5.536 $\pm$ 0.113 & 1.081 $\pm$ 0.056 & C4\\
 & 0.006 $\pm$ 0.006 & 2.217 $\pm$ 0.155 & -7.593 $\pm$ 0.155 & 1.512 $\pm$ 1.079 & C3\\
 & 0.065 $\pm$ 0.020 & 2.721 $\pm$ 0.181 & -12.818 $\pm$ 0.181 & 1.785 $\pm$ 0.170 & C1\\
\hline
2004.63 & 0.325 $\pm$ 0.028 & 0.000 $\pm$ 0.051 & 0.000 $\pm$ 0.029 & 0.041 $\pm$ 0.007 & CS\\
 & 0.694 $\pm$ 0.042 & 0.072 $\pm$ 0.051 & -0.294 $\pm$ 0.030 & 0.073 $\pm$ 0.003 & Cg\\
 & 1.189 $\pm$ 0.061 & 0.187 $\pm$ 0.053 & -0.651 $\pm$ 0.032 & 0.138 $\pm$ 0.002 & C12\\
 & 0.602 $\pm$ 0.043 & 0.322 $\pm$ 0.052 & -0.857 $\pm$ 0.031 & 0.107 $\pm$ 0.005 & C11\\
 & 0.122 $\pm$ 0.023 & 0.614 $\pm$ 0.074 & -1.561 $\pm$ 0.061 & 0.533 $\pm$ 0.037 & C10\\
 & 0.057 $\pm$ 0.016 & 1.091 $\pm$ 0.079 & -2.354 $\pm$ 0.067 & 0.602 $\pm$ 0.084 & C9\\
 & 0.059 $\pm$ 0.016 & 1.082 $\pm$ 0.064 & -3.033 $\pm$ 0.048 & 0.387 $\pm$ 0.068 & C8\\
 & 0.075 $\pm$ 0.013 & 0.897 $\pm$ 0.102 & -3.977 $\pm$ 0.093 & 0.888 $\pm$ 0.038 & C7\\
 & 0.057 $\pm$ 0.012 & 1.516 $\pm$ 0.112 & -5.808 $\pm$ 0.104 & 0.997 $\pm$ 0.057 & C4\\
 & 0.056 $\pm$ 0.021 & 2.724 $\pm$ 0.224 & -13.534 $\pm$ 0.220 & 2.178 $\pm$ 0.319 & C1\\
\hline
2005.22 & 0.315 $\pm$ 0.029 & 0.000 $\pm$ 0.032 & 0.000 $\pm$ 0.030 & 0.029 $\pm$ 0.003 & CS\\
 & 0.584 $\pm$ 0.039 & 0.086 $\pm$ 0.035 & -0.274 $\pm$ 0.033 & 0.133 $\pm$ 0.002 & Cg\\
 & 1.641 $\pm$ 0.068 & 0.247 $\pm$ 0.037 & -0.743 $\pm$ 0.035 & 0.181 $\pm$ 0.001 & C12\\
 & 0.527 $\pm$ 0.039 & 0.371 $\pm$ 0.034 & -0.941 $\pm$ 0.032 & 0.117 $\pm$ 0.002 & C11\\
 & 0.110 $\pm$ 0.018 & 0.682 $\pm$ 0.062 & -1.587 $\pm$ 0.061 & 0.529 $\pm$ 0.016 & C10\\
 & 0.046 $\pm$ 0.012 & 1.262 $\pm$ 0.062 & -2.346 $\pm$ 0.061 & 0.529 $\pm$ 0.037 & C9\\
 & 0.103 $\pm$ 0.017 & 1.120 $\pm$ 0.064 & -3.118 $\pm$ 0.063 & 0.552 $\pm$ 0.017 & C8\\
 & 0.066 $\pm$ 0.017 & 0.869 $\pm$ 0.106 & -4.220 $\pm$ 0.105 & 1.005 $\pm$ 0.065 & C7\\
 & 0.056 $\pm$ 0.014 & 1.482 $\pm$ 0.106 & -5.888 $\pm$ 0.106 & 1.014 $\pm$ 0.061 & C4\\
 & 0.040 $\pm$ 0.016 & 2.553 $\pm$ 0.181 & -13.623 $\pm$ 0.181 & 1.782 $\pm$ 0.262 & C1\\
\hline
2005.56 & 0.400 $\pm$ 0.031 & 0.000 $\pm$ 0.031 & 0.000 $\pm$ 0.033 & 0.075 $\pm$ 0.002 & CS\\
 & 0.533 $\pm$ 0.035 & 0.092 $\pm$ 0.032 & -0.299 $\pm$ 0.034 & 0.110 $\pm$ 0.002 & Cg\\
 & 1.527 $\pm$ 0.060 & 0.275 $\pm$ 0.036 & -0.800 $\pm$ 0.038 & 0.193 $\pm$ 0.001 & C12\\
 & 0.802 $\pm$ 0.043 & 0.378 $\pm$ 0.033 & -0.987 $\pm$ 0.035 & 0.136 $\pm$ 0.001 & C11\\
 & 0.086 $\pm$ 0.014 & 0.644 $\pm$ 0.052 & -1.563 $\pm$ 0.053 & 0.422 $\pm$ 0.015 & C10\\
 & 0.068 $\pm$ 0.015 & 1.228 $\pm$ 0.063 & -2.332 $\pm$ 0.064 & 0.554 $\pm$ 0.029 & C9\\
 & 0.110 $\pm$ 0.019 & 1.140 $\pm$ 0.058 & -3.211 $\pm$ 0.059 & 0.494 $\pm$ 0.017 & C8\\
 & 0.067 $\pm$ 0.015 & 0.906 $\pm$ 0.111 & -4.315 $\pm$ 0.112 & 1.072 $\pm$ 0.053 & C7\\
 & 0.062 $\pm$ 0.015 & 1.585 $\pm$ 0.124 & -5.965 $\pm$ 0.124 & 1.199 $\pm$ 0.070 & C4\\
 & 0.058 $\pm$ 0.019 & 2.721 $\pm$ 0.220 & -13.583 $\pm$ 0.220 & 2.179 $\pm$ 0.232 & C1\\
\hline
2006.32 & 0.493 $\pm$ 0.039 & 0.000 $\pm$ 0.034 & 0.000 $\pm$ 0.032 & 0.055 $\pm$ 0.003 & CS\\
 & 0.620 $\pm$ 0.044 & 0.076 $\pm$ 0.034 & -0.220 $\pm$ 0.031 & 0.027 $\pm$ 0.002 & Cg\\
 & 0.315 $\pm$ 0.041 & 0.213 $\pm$ 0.038 & -0.641 $\pm$ 0.036 & 0.180 $\pm$ 0.008 & C13\\
 & 0.891 $\pm$ 0.069 & 0.370 $\pm$ 0.041 & -0.980 $\pm$ 0.039 & 0.227 $\pm$ 0.003 & C12\\
 & 0.751 $\pm$ 0.064 & 0.461 $\pm$ 0.039 & -1.172 $\pm$ 0.037 & 0.198 $\pm$ 0.004 & C11\\
 & 0.086 $\pm$ 0.024 & 1.248 $\pm$ 0.092 & -2.495 $\pm$ 0.091 & 0.852 $\pm$ 0.071 & C9\\
 & 0.106 $\pm$ 0.020 & 1.174 $\pm$ 0.061 & -3.389 $\pm$ 0.060 & 0.512 $\pm$ 0.022 & C8\\
 & 0.054 $\pm$ 0.016 & 0.931 $\pm$ 0.110 & -4.505 $\pm$ 0.109 & 1.049 $\pm$ 0.100 & C7\\
 & 0.052 $\pm$ 0.016 & 1.649 $\pm$ 0.134 & -6.314 $\pm$ 0.133 & 1.298 $\pm$ 0.130 & C4\\
 & 0.035 $\pm$ 0.011 & 2.680 $\pm$ 0.149 & -13.720 $\pm$ 0.149 & 1.455 $\pm$ 0.156 & C1\\
\hline
2007.17 & 0.719 $\pm$ 0.056 & 0.000 $\pm$ 0.036 & 0.000 $\pm$ 0.035 & 0.092 $\pm$ 0.003 & CS\\
 & 0.549 $\pm$ 0.049 & 0.067 $\pm$ 0.035 & -0.211 $\pm$ 0.034 & 0.049 $\pm$ 0.004 & Cg\\
 & 0.218 $\pm$ 0.031 & 0.258 $\pm$ 0.037 & -0.660 $\pm$ 0.036 & 0.125 $\pm$ 0.011 & C13\\
 & 0.667 $\pm$ 0.054 & 0.370 $\pm$ 0.040 & -1.077 $\pm$ 0.039 & 0.194 $\pm$ 0.004 & C12\\
 & 0.647 $\pm$ 0.054 & 0.552 $\pm$ 0.043 & -1.278 $\pm$ 0.042 & 0.254 $\pm$ 0.004 & C11\\
 & 0.052 $\pm$ 0.015 & 1.415 $\pm$ 0.059 & -2.592 $\pm$ 0.059 & 0.481 $\pm$ 0.057 & C9\\
 & 0.128 $\pm$ 0.023 & 1.199 $\pm$ 0.068 & -3.439 $\pm$ 0.067 & 0.580 $\pm$ 0.024 & C8\\
 & 0.044 $\pm$ 0.011 & 1.059 $\pm$ 0.106 & -4.866 $\pm$ 0.105 & 0.999 $\pm$ 0.064 & C7\\
 & 0.037 $\pm$ 0.010 & 1.726 $\pm$ 0.112 & -6.441 $\pm$ 0.112 & 1.065 $\pm$ 0.084 & C4\\
 & 0.043 $\pm$ 0.018 & 2.626 $\pm$ 0.247 & -14.197 $\pm$ 0.247 & 2.444 $\pm$ 0.400 & C1\\
\hline
2007.50 & 0.783 $\pm$ 0.047 & 0.000 $\pm$ 0.033 & 0.000 $\pm$ 0.033 & 0.052 $\pm$ 0.002 & CS\\
 & 0.587 $\pm$ 0.041 & 0.063 $\pm$ 0.034 & -0.221 $\pm$ 0.034 & 0.090 $\pm$ 0.002 & Cg\\
 & 0.303 $\pm$ 0.035 & 0.264 $\pm$ 0.037 & -0.683 $\pm$ 0.037 & 0.172 $\pm$ 0.006 & C13\\
 & 0.651 $\pm$ 0.051 & 0.388 $\pm$ 0.041 & -1.160 $\pm$ 0.041 & 0.240 $\pm$ 0.003 & C12\\
 & 0.616 $\pm$ 0.049 & 0.592 $\pm$ 0.043 & -1.380 $\pm$ 0.043 & 0.277 $\pm$ 0.003 & C11\\
 & 0.072 $\pm$ 0.018 & 1.398 $\pm$ 0.074 & -2.682 $\pm$ 0.074 & 0.661 $\pm$ 0.046 & C9\\
 & 0.120 $\pm$ 0.022 & 1.255 $\pm$ 0.061 & -3.561 $\pm$ 0.062 & 0.519 $\pm$ 0.022 & C8\\
 & 0.041 $\pm$ 0.011 & 0.995 $\pm$ 0.109 & -4.667 $\pm$ 0.110 & 1.045 $\pm$ 0.082 & C7\\
 & 0.044 $\pm$ 0.013 & 1.633 $\pm$ 0.133 & -6.393 $\pm$ 0.133 & 1.285 $\pm$ 0.111 & C4\\
 & 0.039 $\pm$ 0.016 & 2.657 $\pm$ 0.241 & -14.479 $\pm$ 0.241 & 2.384 $\pm$ 0.366 & C1\\
\hline
2008.58 & 0.699 $\pm$ 0.051 & 0.000 $\pm$ 0.032 & 0.000 $\pm$ 0.033 & 0.067 $\pm$ 0.002 & CS\\
 & 0.540 $\pm$ 0.044 & 0.062 $\pm$ 0.032 & -0.218 $\pm$ 0.033 & 0.068 $\pm$ 0.003 & Cg\\
 & 0.315 $\pm$ 0.028 & 0.252 $\pm$ 0.035 & -0.703 $\pm$ 0.035 & 0.147 $\pm$ 0.003 & C14\\
 & 0.461 $\pm$ 0.065 & 0.383 $\pm$ 0.036 & -0.965 $\pm$ 0.037 & 0.182 $\pm$ 0.009 & C13\\
 & 0.707 $\pm$ 0.081 & 0.629 $\pm$ 0.052 & -1.580 $\pm$ 0.052 & 0.415 $\pm$ 0.007 & C11\\
 & 0.046 $\pm$ 0.013 & 1.455 $\pm$ 0.064 & -2.837 $\pm$ 0.064 & 0.557 $\pm$ 0.047 & C9\\
 & 0.115 $\pm$ 0.020 & 1.306 $\pm$ 0.074 & -3.810 $\pm$ 0.074 & 0.668 $\pm$ 0.022 & C8\\
 & 0.029 $\pm$ 0.014 & 1.020 $\pm$ 0.108 & -5.262 $\pm$ 0.108 & 1.032 $\pm$ 0.215 & C7\\
 & 0.035 $\pm$ 0.016 & 1.717 $\pm$ 0.138 & -6.687 $\pm$ 0.139 & 1.348 $\pm$ 0.270 & C4\\
 & 0.036 $\pm$ 0.019 & 2.453 $\pm$ 0.228 & -14.731 $\pm$ 0.228 & 2.262 $\pm$ 0.551 & C1\\
\hline
2009.02 & 0.778 $\pm$ 0.055 & 0.000 $\pm$ 0.036 & 0.000 $\pm$ 0.033 & 0.058 $\pm$ 0.003 & CS\\
 & 0.616 $\pm$ 0.049 & 0.047 $\pm$ 0.036 & -0.191 $\pm$ 0.033 & 0.062 $\pm$ 0.003 & Cg\\
 & 0.333 $\pm$ 0.049 & 0.199 $\pm$ 0.041 & -0.683 $\pm$ 0.038 & 0.210 $\pm$ 0.011 & C14\\
 & 0.529 $\pm$ 0.061 & 0.405 $\pm$ 0.041 & -1.011 $\pm$ 0.038 & 0.210 $\pm$ 0.007 & C13\\
 & 0.551 $\pm$ 0.063 & 0.667 $\pm$ 0.059 & -1.689 $\pm$ 0.057 & 0.469 $\pm$ 0.009 & C11\\
 & 0.036 $\pm$ 0.009 & 1.450 $\pm$ 0.056 & -2.972 $\pm$ 0.054 & 0.433 $\pm$ 0.039 & C9\\
 & 0.103 $\pm$ 0.016 & 1.274 $\pm$ 0.080 & -3.902 $\pm$ 0.078 & 0.716 $\pm$ 0.020 & C8\\
 & 0.039 $\pm$ 0.015 & 1.373 $\pm$ 0.155 & -5.796 $\pm$ 0.155 & 1.513 $\pm$ 0.232 & C7\\
 & 0.013 $\pm$ 0.007 & 1.981 $\pm$ 0.111 & -7.423 $\pm$ 0.110 & 1.051 $\pm$ 0.260 & C4\\
 & 0.032 $\pm$ 0.019 & 2.707 $\pm$ 0.254 & -14.860 $\pm$ 0.254 & 2.520 $\pm$ 0.751 & C1\\
\hline
2009.41 & 0.906 $\pm$ 0.069 & 0.000 $\pm$ 0.034 & 0.000 $\pm$ 0.032 & 0.075 $\pm$ 0.002 & CS\\
 & 1.186 $\pm$ 0.079 & 0.051 $\pm$ 0.033 & -0.179 $\pm$ 0.032 & 0.045 $\pm$ 0.002 & Cg\\
 & 0.314 $\pm$ 0.041 & 0.221 $\pm$ 0.037 & -0.740 $\pm$ 0.036 & 0.178 $\pm$ 0.008 & C14\\
 & 0.583 $\pm$ 0.056 & 0.425 $\pm$ 0.040 & -1.087 $\pm$ 0.038 & 0.223 $\pm$ 0.004 & C13\\
 & 0.463 $\pm$ 0.032 & 0.733 $\pm$ 0.060 & -1.800 $\pm$ 0.059 & 0.506 $\pm$ 0.003 & C11\\
 & 0.032 $\pm$ 0.008 & 1.424 $\pm$ 0.051 & -3.028 $\pm$ 0.050 & 0.387 $\pm$ 0.036 & C9\\
 & 0.105 $\pm$ 0.016 & 1.329 $\pm$ 0.084 & -3.977 $\pm$ 0.084 & 0.777 $\pm$ 0.019 & C8\\
 & 0.035 $\pm$ 0.013 & 1.203 $\pm$ 0.138 & -5.820 $\pm$ 0.137 & 1.336 $\pm$ 0.181 & C7\\
 & 0.021 $\pm$ 0.011 & 1.994 $\pm$ 0.176 & -7.459 $\pm$ 0.176 & 1.728 $\pm$ 0.448 & C4\\
 & 0.032 $\pm$ 0.018 & 2.493 $\pm$ 0.243 & -15.049 $\pm$ 0.243 & 2.406 $\pm$ 0.637 & C1\\
\hline
2010.04 & 0.479 $\pm$ 0.038 & 0.000 $\pm$ 0.038 & 0.000 $\pm$ 0.034 & 0.047 $\pm$ 0.004 & CS\\
 & 2.109 $\pm$ 0.076 & 0.079 $\pm$ 0.039 & -0.209 $\pm$ 0.034 & 0.063 $\pm$ 0.001 & Cg\\
 & 0.337 $\pm$ 0.030 & 0.126 $\pm$ 0.038 & -0.389 $\pm$ 0.034 & 0.035 $\pm$ 0.005 & C15\\
 & 0.279 $\pm$ 0.031 & 0.319 $\pm$ 0.041 & -0.928 $\pm$ 0.038 & 0.164 $\pm$ 0.007 & C14\\
 & 0.391 $\pm$ 0.037 & 0.511 $\pm$ 0.048 & -1.279 $\pm$ 0.045 & 0.295 $\pm$ 0.006 & C13\\
 & 0.292 $\pm$ 0.033 & 0.889 $\pm$ 0.073 & -2.064 $\pm$ 0.071 & 0.621 $\pm$ 0.011 & C11\\
 & 0.106 $\pm$ 0.018 & 1.397 $\pm$ 0.097 & -4.032 $\pm$ 0.095 & 0.893 $\pm$ 0.032 & C8\\
 & 0.029 $\pm$ 0.011 & 1.196 $\pm$ 0.127 & -6.013 $\pm$ 0.126 & 1.214 $\pm$ 0.190 & C7\\
 & 0.021 $\pm$ 0.011 & 2.142 $\pm$ 0.193 & -7.874 $\pm$ 0.192 & 1.888 $\pm$ 0.487 & C4\\
 & 0.041 $\pm$ 0.035 & 2.501 $\pm$ 0.418 & -15.792 $\pm$ 0.417 & 4.159 $\pm$ 2.074 & C1\\
\hline
2010.74 & 1.167 $\pm$ 0.067 & 0.000 $\pm$ 0.031 & 0.000 $\pm$ 0.032 & 0.055 $\pm$ 0.001 & CS\\
 & 2.185 $\pm$ 0.091 & 0.097 $\pm$ 0.032 & -0.180 $\pm$ 0.033 & 0.089 $\pm$ 0.001 & Cg\\
 & 0.213 $\pm$ 0.037 & 0.182 $\pm$ 0.035 & -0.534 $\pm$ 0.036 & 0.174 $\pm$ 0.012 & C15\\
 & 0.302 $\pm$ 0.044 & 0.398 $\pm$ 0.036 & -1.016 $\pm$ 0.037 & 0.200 $\pm$ 0.009 & C14\\
 & 0.270 $\pm$ 0.042 & 0.570 $\pm$ 0.043 & -1.354 $\pm$ 0.043 & 0.297 $\pm$ 0.010 & C13\\
 & 0.214 $\pm$ 0.039 & 1.053 $\pm$ 0.080 & -2.185 $\pm$ 0.080 & 0.737 $\pm$ 0.025 & C11\\
 & 0.101 $\pm$ 0.019 & 1.400 $\pm$ 0.108 & -4.068 $\pm$ 0.108 & 1.035 $\pm$ 0.036 & C8\\
 & 0.023 $\pm$ 0.009 & 1.210 $\pm$ 0.108 & -5.870 $\pm$ 0.108 & 1.038 $\pm$ 0.161 & C7\\
 & 0.023 $\pm$ 0.013 & 2.226 $\pm$ 0.175 & -7.882 $\pm$ 0.175 & 1.725 $\pm$ 0.474 & C4\\
 & 0.021 $\pm$ 0.013 & 2.715 $\pm$ 0.199 & -15.223 $\pm$ 0.199 & 1.967 $\pm$ 0.648 & C1\\
\hline
2011.40 & 0.601 $\pm$ 0.044 & 0.000 $\pm$ 0.031 & 0.000 $\pm$ 0.033 & 0.058 $\pm$ 0.002 & CS\\
 & 1.676 $\pm$ 0.073 & 0.090 $\pm$ 0.032 & -0.210 $\pm$ 0.034 & 0.087 $\pm$ 0.001 & Cg\\
 & 0.649 $\pm$ 0.045 & 0.269 $\pm$ 0.031 & -0.376 $\pm$ 0.033 & 0.045 $\pm$ 0.002 & C16\\
 & 0.362 $\pm$ 0.034 & 0.191 $\pm$ 0.033 & -0.640 $\pm$ 0.035 & 0.116 $\pm$ 0.004 & C15\\
 & 0.190 $\pm$ 0.022 & 0.465 $\pm$ 0.036 & -1.137 $\pm$ 0.038 & 0.187 $\pm$ 0.006 & C14\\
 & 0.238 $\pm$ 0.025 & 0.641 $\pm$ 0.043 & -1.473 $\pm$ 0.045 & 0.308 $\pm$ 0.005 & C13\\
 & 0.182 $\pm$ 0.026 & 1.152 $\pm$ 0.093 & -2.372 $\pm$ 0.094 & 0.880 $\pm$ 0.019 & C11\\
 & 0.090 $\pm$ 0.019 & 1.394 $\pm$ 0.119 & -4.201 $\pm$ 0.119 & 1.146 $\pm$ 0.053 & C8\\
 & 0.029 $\pm$ 0.013 & 1.408 $\pm$ 0.158 & -6.456 $\pm$ 0.159 & 1.555 $\pm$ 0.274 & C7\\
 & 0.011 $\pm$ 0.011 & 2.548 $\pm$ 0.206 & -8.944 $\pm$ 0.206 & 2.036 $\pm$ 1.275 & C4\\
 & 0.029 $\pm$ 0.026 & 2.422 $\pm$ 0.360 & -15.656 $\pm$ 0.360 & 3.586 $\pm$ 1.971 & C1\\
\hline
2012.04 & 0.469 $\pm$ 0.038 & 0.000 $\pm$ 0.036 & 0.000 $\pm$ 0.036 & 0.020 $\pm$ 0.004 & CS\\
 & 1.595 $\pm$ 0.069 & 0.079 $\pm$ 0.037 & -0.177 $\pm$ 0.037 & 0.094 $\pm$ 0.001 & Cg\\
 & 0.980 $\pm$ 0.055 & 0.252 $\pm$ 0.038 & -0.510 $\pm$ 0.038 & 0.136 $\pm$ 0.002 & C16\\
 & 0.343 $\pm$ 0.032 & 0.263 $\pm$ 0.037 & -0.802 $\pm$ 0.038 & 0.115 $\pm$ 0.005 & C15\\
 & 0.234 $\pm$ 0.029 & 0.575 $\pm$ 0.046 & -1.373 $\pm$ 0.046 & 0.286 $\pm$ 0.010 & C14\\
 & 0.084 $\pm$ 0.018 & 0.786 $\pm$ 0.047 & -1.750 $\pm$ 0.048 & 0.313 $\pm$ 0.028 & C13\\
 & 0.142 $\pm$ 0.018 & 1.309 $\pm$ 0.087 & -2.678 $\pm$ 0.087 & 0.796 $\pm$ 0.015 & C11\\
 & 0.071 $\pm$ 0.020 & 1.441 $\pm$ 0.115 & -4.279 $\pm$ 0.115 & 1.093 $\pm$ 0.092 & C8\\
 & 0.024 $\pm$ 0.009 & 1.398 $\pm$ 0.162 & -6.179 $\pm$ 0.163 & 1.585 $\pm$ 0.233 & C7\\
 & 0.014 $\pm$ 0.008 & 2.310 $\pm$ 0.177 & -8.538 $\pm$ 0.177 & 1.734 $\pm$ 0.487 & C4\\
 & 0.031 $\pm$ 0.035 & 2.124 $\pm$ 0.455 & -16.386 $\pm$ 0.455 & 4.534 $\pm$ 3.414 & C1\\
\hline
2012.59 & 0.338 $\pm$ 0.029 & 0.000 $\pm$ 0.034 & 0.000 $\pm$ 0.034 & 0.050 $\pm$ 0.003 & CS\\
 & 1.652 $\pm$ 0.063 & 0.106 $\pm$ 0.036 & -0.199 $\pm$ 0.035 & 0.123 $\pm$ 0.001 & Cg\\
 & 0.771 $\pm$ 0.036 & 0.303 $\pm$ 0.038 & -0.632 $\pm$ 0.037 & 0.171 $\pm$ 0.001 & C16\\
 & 0.387 $\pm$ 0.025 & 0.309 $\pm$ 0.036 & -0.886 $\pm$ 0.035 & 0.124 $\pm$ 0.002 & C15\\
 & 0.199 $\pm$ 0.028 & 0.619 $\pm$ 0.045 & -1.509 $\pm$ 0.045 & 0.303 $\pm$ 0.011 & C14\\
 & 0.078 $\pm$ 0.018 & 0.883 $\pm$ 0.062 & -1.872 $\pm$ 0.062 & 0.523 $\pm$ 0.035 & C13\\
 & 0.110 $\pm$ 0.021 & 1.515 $\pm$ 0.071 & -2.864 $\pm$ 0.071 & 0.628 $\pm$ 0.029 & C11\\
 & 0.070 $\pm$ 0.020 & 1.310 $\pm$ 0.117 & -3.705 $\pm$ 0.117 & 1.119 $\pm$ 0.097 & C9\\
 & 0.041 $\pm$ 0.016 & 1.549 $\pm$ 0.125 & -4.966 $\pm$ 0.125 & 1.200 $\pm$ 0.187 & C8\\
 & 0.028 $\pm$ 0.019 & 1.811 $\pm$ 0.274 & -7.896 $\pm$ 0.274 & 2.716 $\pm$ 0.959 & C4\\
 & 0.028 $\pm$ 0.034 & 2.225 $\pm$ 0.486 & -16.406 $\pm$ 0.486 & 4.844 $\pm$ 3.991 & C1\\
\hline
2012.84 & 0.492 $\pm$ 0.032 & 0.000 $\pm$ 0.034 & 0.000 $\pm$ 0.032 & 0.059 $\pm$ 0.002 & CS\\
 & 1.328 $\pm$ 0.053 & 0.087 $\pm$ 0.035 & -0.177 $\pm$ 0.033 & 0.095 $\pm$ 0.001 & Cg\\
 & 0.606 $\pm$ 0.040 & 0.289 $\pm$ 0.037 & -0.642 $\pm$ 0.035 & 0.143 $\pm$ 0.002 & C16\\
 & 0.349 $\pm$ 0.031 & 0.323 $\pm$ 0.036 & -0.904 $\pm$ 0.034 & 0.119 $\pm$ 0.004 & C15\\
 & 0.167 $\pm$ 0.022 & 0.607 $\pm$ 0.044 & -1.529 $\pm$ 0.043 & 0.288 $\pm$ 0.009 & C14\\
 & 0.066 $\pm$ 0.012 & 0.926 $\pm$ 0.069 & -1.952 $\pm$ 0.068 & 0.600 $\pm$ 0.024 & C13\\
 & 0.096 $\pm$ 0.018 & 1.521 $\pm$ 0.068 & -2.890 $\pm$ 0.067 & 0.586 $\pm$ 0.023 & C11\\
 & 0.062 $\pm$ 0.015 & 1.332 $\pm$ 0.099 & -3.646 $\pm$ 0.098 & 0.929 $\pm$ 0.062 & C9\\
 & 0.038 $\pm$ 0.009 & 1.462 $\pm$ 0.109 & -4.772 $\pm$ 0.109 & 1.041 $\pm$ 0.059 & C8\\
 & 0.015 $\pm$ 0.007 & 1.382 $\pm$ 0.144 & -6.553 $\pm$ 0.143 & 1.395 $\pm$ 0.264 & C7\\
 & 0.015 $\pm$ 0.008 & 2.377 $\pm$ 0.180 & -8.840 $\pm$ 0.179 & 1.765 $\pm$ 0.444 & C4\\
 & 0.028 $\pm$ 0.030 & 2.331 $\pm$ 0.455 & -16.564 $\pm$ 0.455 & 4.539 $\pm$ 3.146 & C1\\
\hline
2013.06 & 0.412 $\pm$ 0.029 & 0.000 $\pm$ 0.036 & 0.000 $\pm$ 0.034 & 0.037 $\pm$ 0.003 & CS\\
 & 1.535 $\pm$ 0.056 & 0.085 $\pm$ 0.037 & -0.203 $\pm$ 0.035 & 0.098 $\pm$ 0.001 & Cg\\
 & 0.562 $\pm$ 0.034 & 0.300 $\pm$ 0.038 & -0.700 $\pm$ 0.036 & 0.139 $\pm$ 0.002 & C16\\
 & 0.362 $\pm$ 0.027 & 0.346 $\pm$ 0.039 & -0.976 $\pm$ 0.037 & 0.168 $\pm$ 0.003 & C15\\
 & 0.153 $\pm$ 0.021 & 0.635 $\pm$ 0.043 & -1.609 $\pm$ 0.041 & 0.247 $\pm$ 0.011 & C14\\
 & 0.064 $\pm$ 0.015 & 0.948 $\pm$ 0.072 & -2.069 $\pm$ 0.071 & 0.631 $\pm$ 0.041 & C13\\
 & 0.098 $\pm$ 0.018 & 1.552 $\pm$ 0.064 & -2.989 $\pm$ 0.063 & 0.538 $\pm$ 0.023 & C11\\
 & 0.074 $\pm$ 0.014 & 1.359 $\pm$ 0.097 & -3.725 $\pm$ 0.097 & 0.908 $\pm$ 0.037 & C9\\
 & 0.040 $\pm$ 0.011 & 1.499 $\pm$ 0.117 & -4.905 $\pm$ 0.116 & 1.113 $\pm$ 0.093 & C8\\
 & 0.014 $\pm$ 0.006 & 1.421 $\pm$ 0.144 & -6.858 $\pm$ 0.143 & 1.395 $\pm$ 0.214 & C7\\
 & 0.014 $\pm$ 0.007 & 2.285 $\pm$ 0.147 & -8.993 $\pm$ 0.146 & 1.426 $\pm$ 0.301 & C4\\
 & 0.028 $\pm$ 0.025 & 2.251 $\pm$ 0.422 & -16.326 $\pm$ 0.422 & 4.206 $\pm$ 2.293 & C1\\
\hline
\end{longtable}
\end{center}

\begin{table*}
\begin{center}
\caption{The inclination and position angles of the jet axis obtained from the parametric fit on the 15 GHz data.}
\begin{tabular}{@{}c c c}
\hline
\hline
Epoch & $\lambda_0$ & $\iota_0$ \\
\hline
1995.96 & 153.05 $\pm$ 6.38 & 8.20 $\pm$ 1.13 \\ 
1996.05 & 153.67 $\pm$ 4.75 & 10.14 $\pm$ 1.23 \\ 
1996.22 & 154.41 $\pm$ 5.36 & 9.26 $\pm$ 1.25 \\ 
1996.38 & 153.04 $\pm$ 6.33 & 9.04 $\pm$ 1.07 \\ 
1996.57 & 158.42 $\pm$ 7.40 & 9.70 $\pm$ 1.77 \\ 
1996.74 & 152.34 $\pm$ 6.06 & 10.08 $\pm$ 1.14 \\ 
1996.82 & 158.09 $\pm$ 4.30 & 7.17 $\pm$ 0.68 \\ 
1996.93 & 155.77 $\pm$ 5.72 & 7.57 $\pm$ 1.00 \\
1997.66 & 159.22 $\pm$ 3.17 & 7.31 $\pm$ 0.51 \\ 
1998.21 & 159.40 $\pm$ 6.37 & 9.01 $\pm$ 1.19 \\ 
1998.59 & 158.60 $\pm$ 3.93 & 8.02 $\pm$ 0.62 \\ 
1999.01 & 158.89 $\pm$ 6.15 & 7.64 $\pm$ 0.83 \\ 
1999.54 & 155.62 $\pm$ 6.76 & 8.72 $\pm$ 0.97 \\ 
1999.56 & 156.45 $\pm$ 6.93 & 8.73 $\pm$ 1.00 \\ 
1999.57 & 158.41 $\pm$ 4.97 & 9.20 $\pm$ 1.10 \\ 
1999.79 & 156.98 $\pm$ 7.91 & 8.88 $\pm$ 1.17 \\ 
1999.98 & 157.10 $\pm$ 8.01 & 9.63 $\pm$ 1.27 \\ 
2000.01 & 155.37 $\pm$ 6.86 & 9.46 $\pm$ 1.20 \\ 
2000.35 & 154.52 $\pm$ 6.43 & 9.07 $\pm$ 1.07 \\ 
2001.17 & 155.01 $\pm$ 7.84 & 8.72 $\pm$ 1.46 \\ 
2001.38 & 154.38 $\pm$ 7.99 & 8.56 $\pm$ 1.41 \\ 
2001.50 & 154.22 $\pm$ 8.08 & 8.44 $\pm$ 1.40 \\ 
2001.84 & 155.21 $\pm$ 8.11 & 8.32 $\pm$ 1.43 \\ 
2001.97 & 156.44 $\pm$ 7.87 & 8.63 $\pm$ 1.50 \\ 
2002.02 & 155.68 $\pm$ 8.04 & 8.59 $\pm$ 1.42 \\ 
2002.13 & 156.90 $\pm$ 7.55 & 8.63 $\pm$ 1.48 \\ 
2002.45 & 153.07 $\pm$ 5.21 & 8.68 $\pm$ 0.80 \\ 
2003.66 & 161.05 $\pm$ 4.80 & 7.30 $\pm$ 0.66 \\ 
2004.63 & 164.64 $\pm$ 7.66 & 7.85 $\pm$ 0.73 \\ 
2005.22 & 161.71 $\pm$ 6.04 & 8.79 $\pm$ 0.85 \\ 
2005.56 & 161.05 $\pm$ 5.40 & 9.35 $\pm$ 0.88 \\ 
2006.32 & 159.25 $\pm$ 6.51 & 8.07 $\pm$ 0.83 \\ 
2007.17 & 155.18 $\pm$ 6.48 & 8.45 $\pm$ 0.92 \\ 
2007.50 & 155.19 $\pm$ 6.62 & 8.76 $\pm$ 1.02 \\ 
2008.58 & 156.68 $\pm$ 6.33 & 7.95 $\pm$ 0.95 \\ 
2009.02 & 158.33 $\pm$ 7.04 & 8.03 $\pm$ 1.01 \\ 
2009.41 & 160.33 $\pm$ 6.86 & 8.52 $\pm$ 1.06 \\ 
2010.04 & 162.50 $\pm$ 5.65 & 6.70 $\pm$ 0.66 \\ 
2010.74 & 162.49 $\pm$ 5.19 & 7.86 $\pm$ 0.73 \\ 
2011.40 & 160.72 $\pm$ 4.24 & 6.86 $\pm$ 0.54 \\ 
2012.04 & 160.43 $\pm$ 4.97 & 8.52 $\pm$ 0.74 \\ 
2012.59 & 158.35 $\pm$ 5.74 & 9.25 $\pm$ 0.89 \\ 
2012.84 & 156.87 $\pm$ 7.80 & 8.70 $\pm$ 1.10 \\ 
2013.06 & 156.85 $\pm$ 7.37 & 9.43 $\pm$ 1.13 \\ 
\hline
\hline
\end{tabular}
\end{center}
\end{table*}

\end{document}